\documentclass{LMCS}

\def\doi{9(1:10)2013}
\lmcsheading%
{\doi}
{1--19}
{}
{}
{Jul.~\phantom09, 2012}
{Mar.~\phantom06, 2013}
{}

\usepackage{enumerate}
\usepackage{hyperref}

\usepackage{amssymb}
\usepackage{xspace}

\usepackage{dpda-macros}

\begin{document}

\title[Bisimilarity on Basic Process Algebra]{Bisimilarity on Basic Process Algebra
is in 2-ExpTime
\\
(an explicit proof)}

\author[P. Jan\v{c}ar]{Petr Jan\v{c}ar}	
\address{Technical University Ostrava (FEI V\v{S}B-TUO), Czech Rep.}
\email{petr.jancar@vsb.cz}  
\thanks{%
The work was supported 
by the Czech Grant Agency (GA\v{C}R:P202/11/0340)
and partly by the
European Regional Development Fund in the IT4Innovations Centre of
Excellence project (CZ.1.05/1.1.00/02.0070).
}

\keywords{bisimulation equivalence, basic process algebra, complexity}
\ACMCCS{[{\bf Theory of computation}]: Formal languages and automata
  theory---Formalisms---Rewrite systems; Semantics and reasoning---Program
  semantics---Action semantics; Logic---Logic and verification}
\subjclass{F.4.2, F.3.2.}

\begin{abstract}
  \noindent 
Burkart, Caucal, Steffen (1995) showed a procedure deciding
bisimulation equivalence of processes in Basic Process Algebra (BPA),
i.e. of sequential processes generated by context-free grammars.  They
improved the previous decidability result of Christensen, H\"uttel,
Stirling (1992), since their procedure has obviously an elementary
time complexity and the authors claim that a close analysis would
reveal a double exponential upper bound. Here a self-contained direct
proof of the membership in 2-ExpTime is given. This is done via a
Prover-Refuter game which shows that there is an alternating Turing
machine deciding the problem in exponential space.  The proof uses
similar ingredients (size-measures, decompositions, bases) as the
previous proofs, but one new simplifying factor is an explicit
addition of infinite regular strings to the state space.  An auxiliary
claim also shows an explicit exponential upper bound on the
equivalence level of nonbisimilar normed BPA processes.

The importance of clarifying the 2-ExpTime upper bound for BPA
bisimilarity has recently increased due to the  shift of 
the known lower bound from  
PSpace (Srba, 2002) to 
ExpTime (Kiefer, 2012).
\end{abstract}

\maketitle

\section{Introduction}

The classical language equivalence problems in automata theory have
their counterparts in the bisimulation equivalence problems in process
theory. The computational complexity of bisimulation equivalence
is still 
not fully
settled even for fundamental classes, one of them being 
the class of 
Basic Process Algebra (BPA) processes,
i.e. of sequential processes generated by context-free grammars. 
This concrete research topic started with a result by Baeten, Bergstra, Klop~\cite{BBK2}
who showed decidability in the normed BPA case  
 (where each nonterminal of the underlying context-free
 grammar derives some terminal word).
Christensen, H\"uttel, Stirling~\cite{DBLP:journals/iandc/ChristensenHS95} extended the decidability 
result to the whole BPA class, and 
Burkart, Caucal, Steffen~\cite{DBLP:conf/mfcs/BurkartCS95}
(see also~\cite{BCMS})  showed a procedure with an elementary
complexity, claiming that 
 a close analysis would demonstrate
a double exponential upper bound.
We also note that the normed
 case was subsequently shown to be in 
 \PTIME~\cite{Hirshfeld96} (see~\cite{ConLasota10} for
 the most recent improvement of complexity).

Regarding the lower bounds for the (full) BPA problem,
Srba~\cite{DBLP:journals/acta/Srba03} 
showed \PSPACE-hardness,
and Kiefer~\cite{DBLP:journals/corr/abs-1205-7041}
recently shifted this 
to \ExpTime-hardness
(using the \ExpTime-completeness 
of countdown games~\cite{DBLP:journals/lmcs/JurdzinskiSL08});
he thus also strengthened the lower bound results known 
for (visibly) pushdown
processes~\cite{Kucera10},~\cite{DBLP:journals/corr/abs-0901-2068}
and for weak bisimilarity~\cite{DBLP:journals/tcs/Mayr05}.
This was a bit surprising since the
bisimulation equivalence problem for related classes of basic parallel
processes (generated by commutative context-free grammars) and of
one-counter processes were shown \PSPACE-complete~\cite{DBLP:conf/lics/Jancar03}, \cite{BGJ10}.
The mentioned shift of the lower bound is a natural impulse for looking at the complexity
again; 
confirming the upper bound
which has been a bit vaguely stated in the
literature becomes more important.

Here we show a direct self-contained proof of the fact that 
BPA bisimilarity
is indeed in 2-\ExpTime. This is done 
via a Prover-Refuter game which shows that there is an
alternating Turing machine deciding the problem in exponential space.
The proof uses similar ingredients (size-measures, 
decompositions, bases) as the previous proofs, though
in somewhat different forms;
a new factor is an explicit addition 
of infinite regular strings to the state space. On the whole, the
proof confirms the previously claimed upper bound, simplifies several
technical aspects, and
it might also shed some new light on the structural 
decomposition approach
for deciding bisimilarity.
An auxiliary claim also shows an exponential upper bound on
the equivalence level of nonbisimilar normed BPA processes; 
such a bound seems to have been only implicit
in the previous works.

Section~\ref{sec:prelim} recalls the notion of regular strings, defines the bisimilarity
problem for BPA and states the main result. 
Section~\ref{sec:proof} then shows a proof. It recalls
some simple notions
and observations, including the congruence properties and
decompositions, and then a Prover-Refuter game is defined; it will be obvious 
that Refuter has a winning strategy for negative instances.
The above mentioned exponential upper bound on
the equivalence level of nonbisimilar normed BPA processes,
which is used to show 
that Prover has a winning strategy for positive instances,
is highlighted in Section~\ref{sec:eqlevelbounds}.
Section~\ref{sec:addition} adds some further remarks.

\section{Preliminaries}\label{sec:prelim}

\noindent
Let $\Nat=\{0,1,2,\dots\}$. For a finite set $\calC$,
$\card{\calC}$ is the number of elements of $\calC$, and
$\calC^*$ is the set of finite
sequences of elements of $\calC$,
also called \emph{strings} or \emph{words} over $\calC$. By $\varepsilon$ we
denote the empty sequence and by $|w|$ the length of $w\in \calC^*$.
By
$\calC^\omega$ we denote the set of infinite strings over $\calC$, i.e. the
set of
mappings $\Nat\rightarrow\calC$. 
By $uv$ we denote the concatenation of strings $u,v$. 
For technical convenience,
we might write
$uv$ even when $u$ is infinite but then $uv$ is implicitly
identified with $u$. We put $u^0=\varepsilon$ and $u^{i+1}=uu^i$
(where $i\in\Nat$).
By $u^\omega$ we denote the string
$uuu\cdots$;  $u^\omega=u$ when $u$ is infinite,
and 
$\varepsilon^\omega=\varepsilon$.
If $w=uv$ then $u$
is a \emph{prefix} of $w$; if $u$ is finite then
$v$ is a \emph{suffix} of $w$.  

\subsubsection*{Regular strings}

A \emph{regular string} 
over $\calC$ 
is either a finite string (an element of $\calC^*$) 
or an infinite string (an element of $\calC^\omega$)
of the form
$\beta\gamma\gamma\gamma\dots=\beta\gamma^\omega$
where $\beta,\gamma\in\calC^*$ and
$\gamma\neq\varepsilon$.
(Such infinite strings are also called ultimately periodic words.)
We do not consider nonregular strings.

One infinite regular string can have more ``lasso''
presentations, as shown by
the example
\begin{center}
$BAA(BBABBABBA)^\omega=BA(ABB)^\omega$.
\end{center}
The second presentation is the canonical one, since it has the shortest
cycle ($ABB$) and the shortest prefix ($BA$). We now make this
standard notion precise, while also recalling some standard facts
which will be used later.

For $\alpha\in\calC^*$ we put 
$\round(\alpha)=\{\gamma\beta\mid
\beta\gamma=\alpha\}$.

\begin{prop}\label{prop:helpcanon}
If $\beta_1(\gamma_1)^\omega=\beta_2(\gamma_2)^\omega$ 
then
$(\gamma_2)^\omega=(\gamma'_1)^\omega$ for some
$\gamma'_1\in\round(\gamma_1)$.
\end{prop}

\proof
Since $\beta_1\gamma_1\gamma_1\gamma_1\dots=
\beta_2\gamma_2\gamma_2\gamma_2\dots$, we obviously must have
$\gamma_2\gamma_2\gamma_2\dots = \delta\gamma_1\gamma_1\gamma_1\dots$
for a suffix $\delta$ of $\gamma_1$; let $\gamma_1=\delta'\delta$.
Hence
$(\gamma_2)^\omega=\delta(\delta'\delta)^\omega=(\delta\delta')^\omega$.
\qed

\begin{lem}\label{prop:canonform}
Each regular string $\alpha$ has the unique \emph{prefix}
$\alpha_\prefix$
and the unique \emph{cycle}  $\alpha_\cycle$ such that 
$\alpha=\alpha_\prefix(\alpha_\cycle)^\omega$ 
and, moreover,  $\alpha=\beta\gamma^\omega$ implies
$|\beta|\geq |\alpha_\prefix|$
and $|\gamma|\geq |\alpha_\cycle|$ (if $\beta$ is finite).
\end{lem}

\proof
Suppose $\alpha=\beta_1(\gamma_1)^\omega=\beta_2(\gamma_2)^\omega$.
Using Prop.~\ref{prop:helpcanon}, we get
\begin{center}
$\alpha=\beta_1(\gamma_1)^\omega=
\beta_1(\gamma'_2)^\omega=
\beta_2(\gamma_2)^\omega=
\beta_2(\gamma'_1)^\omega$
\end{center}
for some  $\gamma'_2\in\round(\gamma_2)$ and
 $\gamma'_1\in\round(\gamma_1)$. It is thus obvious that 
$\alpha=\beta\gamma^\omega$ where 
$|\beta|=\min\{|\beta_1|,|\beta_2|\}$ 
and $|\gamma|=\min\{|\gamma_1|,|\gamma_2|\}$.
The claim thus follows easily.
\qed

\noindent
We call $\alpha_\prefix(\alpha_\cycle)^\omega$ the 
\emph{canonical presentation of} $\alpha$
 (where $\alpha_\prefix = \alpha$ and 
 $\alpha_\cycle=\varepsilon$ when
 $\alpha$ is finite).
It is useful to note that the (canonical) cycle of a regular string 
is insensitive to any change of a finite prefix, up to swapping:

\begin{prop}\label{prop:corolcanonform}
For any finite $\beta_1$, $\beta_2$
and any (regular) $\alpha$ we have 
$(\beta_2\alpha)_\cycle\in \round((\beta_1\alpha)_\cycle)$.
\end{prop}

\proof
We have
$\beta_1\alpha=(\beta_1\alpha)_\prefix((\beta_1\alpha)_\cycle)^\omega
=\beta_1\alpha_\prefix(\alpha_\cycle)^\omega$; hence
$|(\beta_1\alpha)_\cycle|\leq |\alpha_\cycle|$ (by Lemma~\ref{prop:canonform}).
On the other hand,
$\alpha=\gamma_1((\beta_1\alpha)_\cycle)^\omega$ for some
finite $\gamma_1$, and thus $|\alpha_\cycle|\leq
|(\beta_1\alpha)_\cycle|$; hence
$|(\beta_1\alpha)_\cycle|=|\alpha_\cycle|$.
Similarly $\alpha=\gamma_2((\beta_2\alpha)_\cycle)^\omega$
for some
finite $\gamma_2$, and we deduce 
$|(\beta_1\alpha)_\cycle|=|(\beta_2\alpha)_\cycle|$.
Since $\gamma_1((\beta_1\alpha)_\cycle)^\omega=
\gamma_2((\beta_2\alpha)_\cycle)^\omega$,
by Prop.~\ref{prop:helpcanon} we easily derive that
$(\beta_2\alpha)_\cycle\in \round((\beta_1\alpha)_\cycle)$.
\qed

We will also (implicitly) use the following simple computational fact.

\begin{prop}\label{prop:computecanonical}
There is a polynomial-time algorithm which, given 
finite strings $\beta$ and $\gamma$,
finds the canonical prefix  
$(\beta\gamma^\omega)_\prefix$ and the canonical cycle 
$(\beta\gamma^\omega)_\cycle$.
\end{prop}

\proof
Even a brute-force approach is sufficient here.
We can systematically explore all $3$-part partitions 
$\beta\gamma=\delta_1\delta_2\delta_3$.
For each of them we can check
whether $\delta_1(\delta_2)^\omega=\beta\gamma^\omega$:
for this we must have 
$\delta_3=(\delta_2)^j\delta$, $\delta_2=\delta\delta'$ 
and $(\delta'\delta)^\omega=\gamma^\omega$; the latter holds 
iff $(\delta'\delta)^{|\gamma|}=\gamma^{|\delta'\delta|}$.
\qed

\subsubsection*{BPA processes}

A \emph{BPA system} is 
defined
as a context-free grammar in Greibach
normal form with no starting nonterminal;
it is a tuple $\calG=(\calN,\act,\calR)$
where $\calN$, $\act$, $\calR$ are finite nonempty 
sets of \emph{nonterminals}
(or variables),
$\emph{actions}$ (or terminals), and rewriting \emph{rules},
respectively. The rules in $\calR$ are of the form $A\gt{a}\alpha$ where 
$A\in\calN$, $a\in\act$, $\alpha\in\calN^*$.
For later convenience 
we assume that for each $A\in\calN$ there is at least one rule of the
form $A\gt{a}\alpha$, i.e., 
there are \emph{no dead nonterminals}.
(But there may still be nonterminals
which 
do not derive any terminal word in the classical language sense.)

With each BPA system $\calG=(\calN,\act,\calR)$ we associate
the \emph{labelled transition system} (\emph{LTS})
$\ltsact=(\states,\act,(\gt{a})_{a\in\act})$
where 
$\states$ is the set of all \emph{regular} strings over $\calN$,
which are also called 
\emph{states} or \emph{processes}.
The \emph{transition relations} $\gt{a}\subseteq \states\times\states$
are defined inductively as follows:
if $A\gt{a}\alpha$ is a rule in $\calR$ then 
$A\beta\gt{a}\alpha\beta$ for any regular string $\beta$.
We also define
$\gt{w}$, for $w\in\act^*$, as usual:
$\alpha\gt{\varepsilon}\alpha$\,; if $\alpha\gt{a}\beta$ and 
 $\beta\gt{u}\gamma$ then  $\alpha\gt{au}\gamma$.

\emph{Remark.}
We note that 
$\ltsact$ is generally nondeterministic, since $\calR$ 
can contain rules $A\gt{a}\alpha$ and  $A\gt{a}\beta$ 
where $\alpha\neq\beta$. 
We also note that if $\alpha$ is a finite string and $\alpha\gt{w}\beta$
then $\beta$ is also finite. 
The convenience of including also infinite regular
strings into $\states$ will become clear later. 

\subsubsection*{Bisimilarity problem for BPA}

Given $\calG=(\calN,\act,\calR)$, 
with the associated LTS $\ltsact=(\states,\act,(\gt{a})_{a\in\act})$,
we say that $\calB\subseteq\states\times\states$ 
\emph{covers} 
$(\alpha,\beta)\in \states\times\states$
if for any transition $\alpha\gt{a}\alpha'$  
there is $\beta\gt{a}\beta'$ 
such that 
$(\alpha',\beta')\in \calB$, 
and for any   $\beta\gt{a}\beta'$ there is $\alpha\gt{a}\alpha'$
such that 
$(\alpha',\beta')\in \calB$.
For subsets $\calB,\calB'$ of $\states\times \states$ we say that
$\calB$ \emph{covers} $\calB'$ if $\calB$
covers each $(\alpha,\beta)\in \calB'$.
A set $\calB$ is a \emph{bisimulation} if $\calB$ covers $\calB$.
States $\alpha,\beta$ are \emph{bisimilar},
denoted $\alpha\sim \beta$, if there is a bisimulation
$\calB$ containing $(\alpha,\beta)$.

The problem \textsc{BPA-Bisim} asks, given $\calG$ and two
nonterminals $X,Y$, if $X\sim Y$.
We will prove the next theorem,
assuming a standard encoding of $\calG,X,Y$.

\begin{thm}\label{th:bpaintwoexptime}
\textsc{BPA-Bisim} is in
\textsc{2-\ExpTime}; i.e.,
there is an algorithm which decides
\textsc{BPA-Bisim} and its
time complexity 
is in $O(2^{2^{pol(n)}})$ for a polynomial $pol$. 
\qed
\end{thm}

\section{Proof of Theorem~\ref{th:bpaintwoexptime}}\label{sec:proof} 

In Subsection~\ref{subsec:standardclaims} we 
define some useful technical notions and observe their 
properties. 
These are variants of the ingredients used in the previous works 
like~\cite{DBLP:journals/iandc/ChristensenHS95,
Hirshfeld96,DBLP:conf/mfcs/BurkartCS95}. The extensions to regular
strings are straightforward but we sketch all the proofs, to be
self-contained.
Subsection~\ref{subsec:PRgame} then describes the crux of the
algorithm, formulated as a Prover-Refuter game. 
Soundness (meaning that Prover cannot force a win when $X\not\sim Y$) 
will be
obvious, 
while completeness (Prover can force a win when $X\sim Y$) is shown in
Subsection~\ref{subsec:completeness}; the proof of 
a crucial technical lemma,
related to normed BPA processes, is separated in 
Subsection~\ref{subsec:normedbounds}.

\subsection{Useful notions
and their properties}\label{subsec:standardclaims}

We consider a BPA system $\calG=(\calN,\act,\calR)$,
with the associated labelled transition system $\ltsact=(\states,\act,(\gt{a})_{a\in\act})$.
We put $\sim_0=\states\times \states$, and let 
$\sim_{i+1}\subseteq\states\times \states$ ($i\in\Nat$) be
the set of all pairs covered by $\sim_{i}$.
We note that $\alpha\not\sim_1 \beta$ iff 
$\alpha,\beta$ enable different sets of
actions.

In the next proposition we also use the 
convention that  $\alpha\beta$ and $\alpha^\omega$ are identified 
with $\alpha$
when $\alpha$ is
infinite.

\begin{prop}\label{prop:basiccongruence}
\hfill
\begin{enumerate}[\em(1)]
\item
The relations $\sim$ and $\sim_i$ (for all $i\in\Nat$) are
equivalences.
\item
If $\alpha\sim_{i+1}\beta$ then $\alpha\sim_{i}\beta$ 
(hence $\sim_0\,\supseteq\,
\sim_1\,\supseteq\,\sim_2\,\supseteq\dots$\,).
\item
We have $\alpha\sim\beta$ iff $\forall i\in\Nat:
\alpha\sim_i\beta$.
\item
If $\alpha\sim_i\beta$ and $\gamma\sim_i\delta$ 
then  $\alpha\gamma\sim_i\beta\delta$.
Hence $\sim$ and $\sim_i$ are
congruences w.r.t. concatenation.
\item
If $\alpha\sim_i \gamma\alpha$ and $\gamma\neq\varepsilon$ 
then $\alpha\sim_i \gamma^{\omega}$. 
(Hence $\alpha\sim \gamma\alpha$ implies 
$\alpha\sim\gamma^{\omega}$.)
\end{enumerate}
\end{prop}

\proof
(1) Bisimilarity, 
i.e. the relation $\sim$, can be easily shown to be the
greatest bisimulation, namely the union of all bisimulations; 
the equivalence conditions can be easily checked. 
For relations $\sim_i$, the equivalence conditions can be
easily established 
by induction on $i$.
\\
(2) can be also easily established by induction on $i$.
\\
(3) The inclusion $\bigcap_{i\in\Nat}\sim_i\,\supseteq\, \sim$ 
is trivial.
Since 
$\ltsact$ is image finite, i.e., for each pair $\alpha\in\states$,
$a\in\act$ there are only finitely many $\beta$ such that 
$\alpha\gt{a}\beta$, the set $\bigcap_{i\in\Nat}\sim_i$ can be easily
checked to be a bisimulation; 
therefore $\bigcap_{i\in\Nat}\sim_i\,\subseteq\, \sim$.
\\
(4) 
Our assumption that there is no dead nonterminal
$A\in\calN$ implies 
$\varepsilon\sim_1 \alpha$ iff $\alpha=\varepsilon$.
By induction on $i$ it is easy to show that
$\alpha\sim_i\alpha'$ implies $\alpha\beta\sim_i\alpha'\beta$ and
$\beta\alpha\sim_i\beta\alpha'$.
\\
(5) By (4) and (1), $\alpha\sim_i \gamma\alpha$ implies
 $\gamma\alpha\sim_i \gamma\gamma\alpha$,  
 $\gamma\gamma\alpha\sim_i \gamma\gamma\gamma\alpha$, $\dots$, and
 thus also $\alpha\sim_i \gamma^i\alpha$.  
The obvious fact $\gamma^i\alpha\sim_i \gamma^\omega$ (when
$\gamma\neq\varepsilon$) thus establishes the claim.
\qed

\emph{Remark.}
The ``no dead nonterminal'' assumption is not crucial for the problem
\textsc{BPA-Bisim}, since we can always add 
a special nonterminal $D$ and a special action $d$, with  
the rules $A\gt{d}A$ for all dead nonterminals $A$ (including $D$),
and finally replace 
the question $X\stackrel{?}{\sim} Y$ with  $XD\stackrel{?}{\sim} YD$.

\smallskip

Points (1)--(3) in
Prop.~\ref{prop:basiccongruence} suggest to define
the \emph{equivalence level}, or the \emph{eq-level},
for each pair of strings:
\begin{center}
$\eqlevel(\alpha,\beta)=k\in\Nat$ if
$\alpha\sim_k\beta$ and $\alpha\not\sim_{k+1}\beta$,
and $\eqlevel(\alpha,\beta)=\omega$ if $\alpha\sim\beta$. 
\end{center}
We stipulate 
$n<\omega$ and 
$\omega+n=\omega-n=\omega+\omega=\omega$ for each $n\in\Nat$.

We observe the following facts.
\begin{prop}\label{prop:basicmatching}
\hfill
\begin{enumerate}[\em(1)]
\item
If $\eqlevel(\alpha,\beta)<\omega$ then either there is 
a transition
$\alpha\gt{a}\alpha'$
such that 
for any 
$\beta\gt{a}\beta'$
we have
$\eqlevel(\alpha',\beta')<\eqlevel(\alpha,\beta)$,
or there is a
transition $\beta\gt{a}\beta'$
 such that 
for any 
$\alpha\gt{a}\alpha'$
we have
$\eqlevel(\alpha',\beta')<\eqlevel(\alpha,\beta)$.
\item
If $\alpha\gt{a_1}\alpha_1\gt{a_2}\alpha_2\cdots \gt{a_k}\alpha_k$ 
where $a_i\in\act$ (for all $i, 1\leq i\leq k$) and 
$k\leq \eqlevel(\alpha,\beta)$ 
then
there are $\beta_1,\beta_2,\dots, \beta_k$
such that 
$\beta\gt{a_1}\beta_1\gt{a_2}\beta_2\cdots \gt{a_k}\beta_k$ 
and 
\\
$\eqlevel(\alpha_i,\beta_i)\geq \eqlevel(\alpha,\beta)-i$
for $i=1,2,\dots,k$; this implies $\alpha_i\sim\beta_i$ if 
$\alpha\sim\beta$.
\item
If $\eqlevel(\alpha,\alpha')\geq \eqlevel(\alpha,\beta)+1$ then
 $\eqlevel(\alpha,\beta)= \eqlevel(\alpha',\beta)$.
\item
$\eqlevel(\alpha,\beta)\leq \eqlevel(\alpha\gamma,\beta\gamma)$. 
\end{enumerate}
\end{prop}

\proof
The claims easily follow from the definitions of $\sim_i$ and $\sim$.
In Point 2 we can use induction on $k$. For Point 3 it suffices to
note that if $\alpha\sim_{i}\beta$, 
$\alpha\not\sim_{i+1}\beta$, and 
$\alpha\sim_{i+1}\alpha'$ (hence also
$\alpha\sim_{i}\alpha'$) then $\alpha'\sim_{i}\beta$ and 
$\alpha'\not\sim_{i+1}\beta$. 
For Point 4 we note that $\alpha\sim_i\beta$ implies 
 $\alpha\gamma\sim_i\beta\gamma$ by
 Prop~\ref{prop:basiccongruence}(1,4). 
\qed

Now we define the \emph{norm}
as a mapping $\states\rightarrow\Nat\cup\{\omega\}$.

\begin{defi}\label{def:norm}
The \emph{norm} of $\alpha\in \states$ is denoted
by $\|\alpha\|$.
 If there is no $w\in\act^*$ such that  $\alpha\gt{w}\varepsilon$ then 
 we put $\|\alpha\|=\omega$ and say that $\alpha$ is \emph{unnormed};
 otherwise $\alpha$ is \emph{normed} and 
$\|\alpha\|=|w|$ for a shortest $w$ such
that $\alpha\gt{w}\varepsilon$. 

A \emph{path} $\beta_0\gt{a_1}\beta_1\gt{a_2}\beta_2\cdots
\gt{a_k}\beta_k$
in $\ltsact$, where $k\geq 1$ and $a_i\in\act$,
is \emph{norm-reducing} if 
$\|\beta_{i}\|>\|\beta_{i+1}\|$ (and thus necessarily $\|\beta_{i+1}\|=\|\beta_i\|-1$)
 for $i=0,1,\dots,k{-}1$.
\end{defi}

We note that 
$\|\varepsilon\|=0$ and 
$\|\alpha\beta\|=\|\alpha\|+\|\beta\|$. 
We have $\|\alpha\|=\omega$ when
$\alpha$ is infinite.
Now we observe further simple facts.

\begin{prop}\label{prop:simplenorm}
\hfill
\begin{enumerate}[\em(1)]
\item
If $\|\alpha\|\neq\|\beta\|$ then 
$\eqlevel(\alpha,\beta)\leq \min\{\|\alpha\|,\|\beta\|\}$
(and thus $\alpha\not\sim\beta$).
\item
If $U\in\calN$ and $\|U\|=\omega$ then 
$U\sim U\alpha$ for any $\alpha$.
\item
$\eqlevel(\gamma\alpha,\gamma\beta)\geq\|\gamma\|+\eqlevel(\alpha,\beta)$.
\end{enumerate}
\end{prop}

\proof
(1)
Suppose $\|\alpha\|<\|\beta\|$.
Hence $\alpha\gt{u}\varepsilon$ for some $u$ where $|u|=\|\alpha\|$.
If $\eqlevel(\alpha,\beta)\geq \|\alpha\|$ then there is $\beta'$ such
that $\beta\gt{u}\beta'$ and 
$\eqlevel(\varepsilon,\beta')\geq \eqlevel(\alpha,\beta)-\|\alpha\|$
(by Prop.~\ref{prop:basicmatching}(2)).
Since  $\|\beta\|>\|\alpha\|$, we have
$\beta'\neq\varepsilon$, and thus $\eqlevel(\varepsilon,\beta')=0$.
Hence 
$\eqlevel(\alpha,\beta)\leq\|\alpha\|$.

(2)
We can easily check that 
the set $\{(\alpha\gamma,\beta\delta) \mid 
\alpha\sim\beta,
\|\alpha\|=\|\beta\|=\omega  \}$ is a bisimulation.

(3) If $\gamma\alpha\sim\gamma\beta$ (which surely holds when
$\|\gamma\|=\omega$) then 
the claim is trivial.
We thus assume $\gamma\alpha\not\sim\gamma\beta$ and
proceed by induction on  $\eqlevel(\gamma\alpha,\gamma\beta)$.
If $\eqlevel(\gamma\alpha,\gamma\beta)=0$ then 
$\gamma=\varepsilon$ (hence $\|\gamma\|=0$) and the claim is trivial.
If $\gamma\neq\varepsilon$ then Prop.~\ref{prop:basicmatching}(1)
implies that there is a transition 
$\gamma\gt{a}\sigma$, where necessarily $\|\sigma\|\geq \|\gamma\|-1$,
such that
$\eqlevel(\sigma\alpha,\sigma\beta)< \eqlevel(\gamma\alpha,\gamma\beta)$.
Since $\eqlevel(\sigma\alpha,\sigma\beta)\geq
\|\sigma\|+\eqlevel(\alpha,\beta)$ by the induction hypothesis,
we deduce $\eqlevel(\gamma\alpha,\gamma\beta)\geq
1+\|\sigma\|+\eqlevel(\alpha,\beta)\geq\|\gamma\|+\eqlevel(\alpha,\beta)$.
\qed

\medskip
\noindent
\emph{Convention.}
Prop.~\ref{prop:simplenorm}(2) 
allows us to remove the 
suffix after the first occurrence of an unnormed nonterminal in any
string, without changing its bisimulation equivalence class.
We thus further implicitly assume that the considered strings 
are of the forms $\alpha$, $\alpha U$, 
or $\beta\gamma^\omega$ 
where $\alpha, \beta, \gamma\in\calN^*$ are normed and 
$U\in\calN$ is unnormed.
We still might write, e.g.,
$\gamma\beta$ or $\gamma^\omega$ even if
$\|\gamma\|=\omega$ but such strings are implicitly identified
with the appropriate prefix of $\gamma$.

\medskip

It will be useful to use the norm when measuring the size of 
string presentations:

\begin{defi}\label{def:basicsizes}
Given $\calG=(\calN,\act,\calR)$,
the function $\size:\states\cup(\states\times\states)\rightarrow\Nat$
is defined as follows.
\begin{iteMize}{$\bullet$}
\item
For a finite string
$\alpha$ we put $\size(\alpha)=\|\alpha'\|$ where $\alpha'$
is the longest normed
prefix of $\alpha$. 
(Thus $\size(\alpha U)=\size(\alpha)=\|\alpha\|$ when
$\alpha$ is normed and
$U$ is unnormed.)
\item
For an infinite regular string $\alpha$, containing
no unnormed nonterminal, 
we put  $\size(\alpha)=\|\alpha_\prefix \alpha_\cycle\|$
(where $\alpha_\prefix(\alpha_\cycle)^\omega$
is the canonical
presentation of $\alpha$). 
\item
For a pair $(\alpha,\beta)$ we put 
$\size(\alpha,\beta)=\max\,\{\size(\alpha),\size(\beta)\}$.
\end{iteMize}
Stipulating $\max\emptyset=0$, we define:
\begin{quote}
$M=\max\,\{\|A\| ; A\in\calN, \|A\|<\omega\}$,
\\
$\Mrhs=
\max\,\{\,\|\alpha\| ;  \|\alpha\|<\omega
\textnormal{ and }\calR
\textnormal{ contains a rule } 
A\gt{a}\alpha\,\}$,
\\
$\Srhs=
\max\,\{\,\size(\alpha)\mid \calR
\textnormal{ contains a rule } 
A\gt{a}\alpha\,\}$.\smallskip 
\end{quote}
\end{defi}

\noindent Hence $M$ is the maximal norm of normed nonterminals, 
and $\Srhs$ is the maximal size of the right-hand sides (rhs) in the rules of $\calG$; 
in particular, $\Srhs$ is greater than or equal to the norm of any normed rhs,
and thus $\Mrhs\leq\Srhs$.

The following fact is also standard; we sketch a proof to be
self-contained.

\begin{prop}\label{prop:norms}
There is a polynomial-time algorithm which, given
$\calG=(\calN,\act,\calR)$, computes $\|A\|$ for each $A\in\calN$,
and also $M,\Mrhs,\Srhs$;
these values are bounded by an exponential function
of the size
of $\calG$.
\end{prop}

\proof
We sketch an algorithm which 
outputs nonterminals in an order 
$A_1,A_2,\dots, A_k$ (for $k=\card{\calN}$) where 
$\|A_1\|\leq \|A_2\|\leq\cdots\leq\|A_k\|$.
Suppose
$A_1,A_2,\dots,A_i$ and their norms 
have been already established ($i=0$ in the beginning).
Construct the set
\begin{center}
$\calD=\{\alpha\mid \alpha\in\{A_1,A_2,\dots,A_i\}^*$
and there is a rule $A\gt{a}\alpha$ for
$A\not\in\{A_1,A_2,\dots,A_i\}\}$.
\end{center}
If $\calD\neq\emptyset$
then put $m=\min\{\,\|\alpha\|\,;\,\alpha\in \calD\,\}$ 
and
define $A_{i+1}$ as a chosen $A\not\in\{A_1,A_2,\dots,A_i\}$
for which there is a rule  $A\gt{a}\alpha$
such that $\alpha\in \calD$ and $\|\alpha\|=m$;
it is obvious that $\|A_{i+1}\|=1+m$.
If $\calD=\emptyset$ then $\|A\|=\omega$ for all  
$A\not\in\{A_1,A_2,\dots,A_i\}$.
The time complexity of the algorithm is obviously polynomial.
The exponential bounds follow by noting that 
$\|A_i\|\leq M_i$ where we put $M_0=0$ and 
$M_{i+1}=1+r\cdot M_i$ for $r=\max\{\,|\alpha| \,;\, \alpha$ is the rhs
of a rule in $\calR\,\}$.
\qed

\emph{Remark.}
The exponential upper bound in the proof is tight:
if we have the rules $A_k\gt{a}A_{k-1}A_{k-1}$, $\dots$,
$A_i\gt{a}A_{i-1}A_{i-1}$, $\dots$, $A_2\gt{a}A_1A_1$,
$A_1\gt{a}\varepsilon$ then $\|A_i\|=2^i-1$.

\smallskip

We now define a crucial notion, 
used in the later Prover-Refuter
game.

\begin{defi}\label{defn:decomp}
A nonempty set 
$\{(\alpha_1,\beta_1)$, $(\alpha_2,\beta_2)$, $\dots$,
$(\alpha_k,\beta_k)\}$ 
is a \emph{decomposition}  of $(\alpha,\beta)$ if 
$\size(\alpha_j,\beta_j)<\size(\alpha,\beta)$ for $j=1,2,\dots,k$,
and  $(\alpha,\beta)$ belongs to
the least congruence (w.r.t. concatenation)
containing all $(\alpha_j,\beta_j)$,
$j=1,2,\dots,k$.
Moreover, if $\alpha_j\sim\beta_j$ for all 
$j=1,2,\dots,k$ then it is a \emph{bisimilar decomposition}.
\end{defi}

\begin{exa}\label{E:3.8}
One decomposition of $(A\alpha,B\beta)$ 
is $\{\,(A\gamma,B), (\alpha,\gamma\beta)\,\}$ when 
both $\size(A\gamma,B)$ and $\size(\alpha,\gamma\beta)$ are less than
 $\size(A\alpha,B\beta)$. Indeed, a \emph{least congruence proof}
 is the sequence  
 $(A\gamma,B)$, $(\alpha,\gamma\beta)$,
$(\beta,\beta)$, $(A\gamma\beta,B\beta)$,
$(A,A)$, $(A\alpha, A\gamma\beta)$, $(A\alpha,B\beta)$
where each pair either is a generator 
($(A\gamma,B)$ or $(\alpha,\gamma\beta)$ in our case) or is deduced
from the previous pairs by using reflexivity, symmetry, transitivity,
and concatenation.
Another decomposition of $(A\alpha,B\beta)$ is 
$\{(\alpha,\gamma\beta), 
(\beta,\delta^\omega),  
(A\gamma\delta^\omega, B\delta^\omega)\}$
if the size conditions are satisfied.
\end{exa}

\begin{prop}\label{prop:basesound}
If $\{(\alpha_j,\beta_j)\mid 1\leq j\leq k\}$
is a decomposition of $(\alpha,\beta)$ 
then 
\begin{center}
$\min\,\{\,\eqlevel(\alpha_j,\beta_j)\mid 1\leq j\leq k\,\}\leq
\eqlevel(\alpha,\beta)$;
\end{center}
if it is a bisimilar decomposition then 
$\alpha\sim\beta$. 
\end{prop}

\proof
Let $(\alpha,\beta)$ belong to the least congruence 
generated by  $\{(\alpha_j,\beta_j)\mid 1\leq j\leq k\}$.
Then there is a least congruence proof $(\gamma_1,\delta_1)$,
$(\gamma_2,\delta_2)$, $\dots$, $(\gamma_m,\delta_m)$
such that $(\gamma_m,\delta_m)=(\alpha,\beta)$, and  
$(\gamma_i,\delta_i)$, for each $i,1\leq i\leq m$, either is 
a generator $(\alpha_j,\beta_j)$, or satisfies $\gamma_i=\delta_i$
(reflexivity), or can be derived from pairs   
$(\gamma_1,\delta_1)$,
$(\gamma_2,\delta_2)$, $\dots$, $(\gamma_{i-1},\delta_{i-1})$
by using
symmetry, transitivity, or concatenation
($\gamma_i=\gamma_{i_1}\gamma_{i_2}$, 
$\delta_i=\delta_{i_1}\delta_{i_2}$ for some $i_1<i$, $i_2<i$).

For any $\ell\in\Nat$, by using the fact that $\sim_\ell$ is 
a congruence w.r.t. concatenation 
(as follows from Prop.~\ref{prop:basiccongruence}(1,4)) we get:
if $\alpha_j\sim_\ell\beta_j$ for all $j,1\leq j\leq k$, then 
$\gamma_i\sim_\ell\delta_i$ for $i=1,2,\dots,m$, and thus
$\alpha\sim_\ell\beta$. 
Hence if  $\alpha_j\sim_\ell\beta_j$ for all $j,1\leq j\leq k$, and all
$\ell\in\Nat$ then 
$\alpha\sim_\ell\beta$
for all $\ell\in\Nat$, and thus $\alpha\sim\beta$
(by Prop.~\ref{prop:basiccongruence}(3)). 
\qed

\subsection{Algorithm deciding \textsc{BPA-Bisim}, based on a Prover-Refuter
game}\label{subsec:PRgame}

We recall that 2-\ExpTime$=$\AExpSpace where ``\textsc{A}''
stands for ``Alternating''~\cite{ChandraKozenStockmeyer}.
For proving Theorem~\ref{th:bpaintwoexptime} it is thus sufficient to
show
an alternating Turing machine working in exponential
space which accepts precisely those $\calG,X,Y$ where $X\not\sim Y$.
The existence of such a machine easily follows from the following
game, 
once we show that Refuter has a winning strategy iff $X\not\sim Y$.

\bigskip

\textsc{Prover} (she) - \textsc{Refuter} (he) \textsc{Game}
\begin{enumerate}[(1)]
\item
A BPA-system $\calG=(\calN,\act,\calR)$ and $X,Y\in\calN$ are given.
\item
A work space of size $2^{pol(size(\calG))}$ is reserved,
where $pol$ is a (sufficiently large) polynomial whose existence will become
clear later.
A part of the work space serves for storing a presentation
of  a \emph{current pair}, initially $(X,Y)$; the rest of the work
space is called the \emph{free work space}.
\item
For $i=1,2,3,\dots$, the following Phase $i$ is performed;
$(\alpha,\beta)$ denotes the current pair:
\begin{enumerate}
\item
If $\alpha\not\sim_1\beta$ then Refuter wins.
If $\alpha,\beta$ are dead (i.e., if they do not enable any action, 
i.e. $\alpha=\beta=\varepsilon$) then Prover wins.
The play finishes in these cases; otherwise it continues with (b).
\item
Prover can decide to show
some (freely chosen) 
pairs and demonstrate  that these pairs
constitute a decomposition of 
$(\alpha,\beta)$.
She is restricted by the free work space when presenting the pairs and 
a least congruence proof. 
(As shown later, it suffices to allow only decompositions with at most
\emph{three} pairs.)
Then Refuter chooses a pair $(\alpha',\beta')$ from the decomposition
and replaces the current pair $(\alpha,\beta)$ with 
 $(\alpha',\beta')$. (Recall that $\size(\alpha',\beta')<\size(\alpha,\beta)$.) 
The play then continues with Phase $i{+}1$.
\item
(Prover has not used the possibility in (b).) 
Refuter chooses a transition $\alpha\gt{a}\alpha'$ or 
$\beta\gt{a}\beta'$. In the first case Prover chooses
some $\beta\gt{a}\beta'$,
in the second case Prover chooses some $\alpha\gt{a}\alpha'$.
If $(\alpha',\beta')$ does not
fit into the space reserved for the current pair 
then Refuter wins; otherwise the current pair $(\alpha,\beta)$ is
replaced with $(\alpha',\beta')$ and
the play continues with Phase $i{+}1$.
\end{enumerate}
\end{enumerate}

\noindent
\emph{Remark.}
A play can be infinite, which can be viewed as a win of Prover. 
To make each play finite, we could add a step counter whose overflow (over
a double exponential bound) would mean that a game configuration has
been repeated and that Prover has won, 
but this is not technically necessary.

\begin{lem}\label{lem:soundness}
(Soundness.) If $X\not\sim Y$ then Refuter has a winning strategy
(even in the game with no space restriction).
\end{lem}
\proof
Assume that $X\not\sim Y$ and Refuter uses the following strategy.
In (b) he always chooses a pair $(\alpha',\beta')$ with the least 
eq-level,
and in (c) he always chooses a transition guaranteeing that 
$\eqlevel(\alpha',\beta')<\eqlevel(\alpha,\beta)$.
Prop.~\ref{prop:basicmatching}(1)
and Prop.~\ref{prop:basesound} show that this is possible and that
$\eqlevel(\alpha',\beta')<\eqlevel(\alpha,\beta)$, or 
$\eqlevel(\alpha',\beta')=\eqlevel(\alpha,\beta)$
and $\size(\alpha',\beta')<\size(\alpha,\beta)$.
Refuter thus must win eventually; he can only benefit from any space
restriction.
\qed

In the next subsection we show the completeness (Prover has a winning
strategy when $X\sim Y$) by which a proof 
of Theorem~\ref{th:bpaintwoexptime} will be finished.

\subsection{Completeness of the Prover-Refuter
game}\label{subsec:completeness}

Our aim is to prove Lemma~\ref{lem:completeness};
a crucial technical fact is captured 
by the next lemma
(assuming a given $\calG=(\calN,\act,\calR)$):

\begin{lem}\label{lem:yielddelta}
If $\alpha_1\not\sim\alpha_2$
and $\alpha_1\beta\sim\alpha_2\beta$ 
then there is $\delta\neq\varepsilon$ such that
$\beta\sim\delta\beta$ 
(and thus $\beta\sim\delta^\omega$)
and
$\size(\delta)\leq (\size(\alpha_1,\alpha_2)+
\card{\calN}^2\cdot\Mrhs+\Srhs)\cdot (1+\Srhs)$.
\qed
\end{lem}
In the lemma we can have $\|\delta\|=\omega$;
in this case $\delta\beta=\delta^\omega=\delta$ (by our convention
after Prop.~\ref{prop:simplenorm}).
We postpone a proof of this lemma,
and a related discussion of normed BPA,
to Subsection~\ref{subsec:normedbounds} and
Section~\ref{sec:eqlevelbounds}.
Now we observe a bound on the possible increase of
the string size in any transition 
in $\ltsact$. 
Roughly speaking, by performing a transition
the canonical cycle either does not change, or is swapped, or 
becomes empty;
the canonical prefix can increase by $\Srhs$ at
most.
\begin{prop}\label{prop:invarstep}
If $\alpha\gt{a}\delta$, 
i.e. $\alpha_\prefix(\alpha_\cycle)^\omega\gt{a}
\delta_\prefix(\delta_\cycle)^\omega$, 
then $\delta_\cycle\in\round(\alpha_\cycle)$ or
$\delta_\cycle=\varepsilon$, hence 
$\size(\delta_\cycle)\leq \size(\alpha_\cycle)$,
and 
$\size(\delta_\prefix)\leq \size(\alpha_\prefix)+\Srhs$.
\end{prop}

\proof
\noindent
We have $\alpha\gt{a}\delta$ due to a rule $A\gt{a}\gamma$,
where $\alpha=A\alpha'$ and  $\delta=\gamma\alpha'$.

If $\|\gamma\|=\omega$ then $\delta=\gamma$ 
(by Convention after Prop.~\ref{prop:simplenorm}), which entails
$\delta_\prefix=\gamma$,
$\delta_\cycle=\varepsilon$, and  
$\size(\delta_\prefix)\leq \Srhs$.

If $\|\gamma\|<\omega$
then (also $\|A\|<\omega$ and) 
$\delta_\cycle\in\round(\alpha_\cycle)$ by
Prop.~\ref{prop:corolcanonform}.
Recalling Lemma~\ref{prop:canonform}, we note that
if $\alpha_\prefix\neq \varepsilon$ then
$\alpha_\prefix=A\alpha'_\prefix$, and 
$\delta_\prefix$ is a prefix of $\gamma\alpha'_\prefix$; this entails 
$\size(\delta_\prefix) < \size(\alpha_\prefix)+\Srhs$.
If $\alpha_\prefix = \varepsilon$ then
$\alpha=(\alpha_c)^\omega=A\beta^\omega$ 
where $\beta\in\round(\alpha_\cycle)$;
hence $\delta=\gamma\beta^\omega$, 
which entails that $\delta_\prefix$ is a prefix of $\gamma$ and thus
$\size(\delta_\prefix) \leq \Srhs$.
\qed

The next technical lemma, Lemma~\ref{lem:detailedcompleteness},
is related to Point 3(b) in the
Prover-Refuter game. It aims to show that if the current pair
is $(A\alpha,B\beta)$ where 
$A\alpha\sim B\beta$ and the presentation size of $(A\alpha,B\beta)$
is bigger than an exponential bound then there is a bisimilar
decomposition of $(A\alpha,B\beta)$, with at most three pairs and with a
least congruence proof of bounded size. 

We handle separately the size of canonical prefixes
and the size of canonical cycles.  
Our convention (after Prop.~\ref{prop:simplenorm})
implies $\size(\alpha_\cycle)=\|\alpha_\cycle\|<\omega$ (including the
case $\alpha_\cycle=\varepsilon$).

\begin{defi}
Given  $\calG=(\calN,\act,\calR)$ and $\bound\in\Nat$,
we say that a (regular) \emph{string} 
$\alpha\in\calN^*\cup \calN^\omega$
\emph{has} an \emph{$\bound$-bounded cycle} if 
$\size(\alpha_\cycle)\leq \bound$.
\end{defi}

In the next lemma,
$\bound$ is an exponential bound w.r.t. the size of $\calG$
(as follows from Prop.~\ref{prop:norms}). The chosen $\bound$ 
and the following analysis are a bit generous,
since we prefer technical simplicity to more detailed upper bounds.

\begin{lem}\label{lem:detailedcompleteness}
Given a BPA system $\calG=(\calN,\act,\calR)$, we put
\begin{center}
$\bound=(2M + \card{\calN}^2\cdot\Mrhs + \Srhs)\cdot (1+\Srhs)$.
\end{center}
If $A\alpha\sim B\beta$, 
both  $A\alpha,B\beta$ have $\bound$-bounded cycles, and
$\size((A\alpha)_\prefix,(B\beta)_\prefix)>2M+\bound$
then 
there is a bisimilar decomposition 
$\{(\alpha_1,\beta_1),(\alpha_2,\beta_2),(\alpha_3,\beta_3)\}$
of $(A\alpha,B\beta)$ 
where all $\alpha_j$, $\beta_j$ ($1\leq j\leq 3$) have
$\bound$-bounded cycles.
\end{lem}

\proof
Let us consider $A\alpha=(A\alpha)_\prefix((A\alpha)_\cycle)^\omega,
B\beta=(B\beta)_\prefix((B\beta)_\cycle)^\omega$ satisfying 
the assumption.
By our convention,  $\alpha=\varepsilon$ if
$\|A\|=\omega$ and $\beta=\varepsilon$ if
$\|B\|=\omega$;
w.l.o.g. we assume $\|A\|\leq \|B\|$.

We recall that
$\size(\alpha,\beta)=\max\{\size(\alpha),\size(\beta)\}$
(by Def.~\ref{def:basicsizes})
and we now show that
\begin{equation}\label{eq:sizedecrease}
\size(\alpha,\beta)<\size(A\alpha,B\beta).
\end{equation}
This is not valid in general, since 
$\size(\alpha)<\size(A\alpha)$ if and only if 
$(A\alpha)_\prefix\neq\varepsilon$;
if $(A\alpha)_\prefix=\varepsilon$ then $A\alpha=((A\alpha)_\cycle)^\omega$, 
$\alpha=((\alpha)_\cycle)^\omega$, and
$\alpha_\cycle\in\round((A\alpha)_\cycle)$, which implies
$\size(\alpha)=\size(A\alpha)$.
In our case we thus have $\size(\alpha)<\size(A\alpha)$ 
or 
$\size(\alpha)=\size(A\alpha)\leq\bound$, and  
$\size(\beta)<\size(B\beta)$ or
$\size(\beta)=\size(B\beta)\leq\bound$.
Since $\size((A\alpha)_\prefix,(B\beta)_\prefix)>2M+\bound$, 
we indeed easily establish~(\ref{eq:sizedecrease}).
Moreover, both $\alpha,\beta$ have $\bound$-bounded cycles as well. 

Now we perform a case analysis 
(showing also some decompositions with even less than three
pairs); recall that we assume $\|A\|\leq \|B\|$.
\begin{enumerate}[(1)]
\item
$\|A\|\leq\|B\|=\omega$; hence $\beta=\varepsilon$, 
$\|A\|<\omega$,
$\alpha\neq\varepsilon$ (since $\size(A\alpha)>2M+\bound>M$),
and $A\alpha\sim B$:

There is a norm-reducing path $A\gt{u}\varepsilon$, 
where $|u|=\|A\|\leq M$;
we have
$A\alpha\gt{u}\alpha$.
By Prop.~\ref{prop:basicmatching}(2) there is $\gamma$ such that
$B\gt{u}\gamma$ and $\alpha\sim\gamma$, and thus also $A\gamma\sim B$
(by Prop.~\ref{prop:basiccongruence}(1,4)); 
recalling
Prop.~\ref{prop:invarstep}, we derive that
$\size(\gamma)\leq \size(B)+M\cdot\Srhs=M\cdot\Srhs$.

We easily check that both $\size(\alpha,\gamma)$ and $\size(A\gamma,B)$
are less than
$\size(A\alpha,B)$, 
and 
that 
$\{(\alpha,\gamma), (A\gamma,B)\}$ is a bisimilar decomposition of 
$(A\alpha,B)$ (as shown by the least congruence proof 
$(\alpha,\gamma)$, $(A,A)$, $(A\alpha,A\gamma)$, 
$(A\gamma,B)$, $(A\alpha,B)$); moreover, all strings in the
decomposition have $\bound$-bounded cycles (which are empty for
$\gamma, A\gamma, B$). 

\item
$\|A\|\leq\|B\|<\omega$ (and $A\alpha\sim B\beta$);
we consider the disjoint cases (a) and (b):

\begin{enumerate}
\item\label{itemdelta}
There is norm-reducing $B\gt{v}\varepsilon$ 
(hence $|v|=\|B\|$, and $B\beta\gt{v}\beta$)
such that 
$A\gt{v}\delta$ 
for some $\delta\neq\varepsilon$
where $\delta\alpha\sim\beta$:

For any norm-reducing $A\gt{u}\varepsilon$ there is surely $\gamma$
such that $B\gt{u}\gamma$ and 
$\alpha\sim\gamma\beta$ 
(since $\|A\|\leq \|B\|$ and $A\alpha\sim B\beta$). 
Since $\size(A)=\|A\|\leq M$ and $\size(B)=\|B\|\leq M$, for (finite)
strings $\gamma, \delta$ we get
\begin{center}
$\size(\gamma)\leq M\cdot(1+\Srhs)\leq \frac{\bound}{2}$, 
$\size(\delta)\leq M\cdot(1+\Srhs)\leq \frac{\bound}{2}$.
\end{center}
Since $\alpha\sim \gamma\beta \sim \gamma\delta\alpha
\sim(\gamma\delta)^\omega$ (recall
Prop.~\ref{prop:basiccongruence}(5)),
and similarly $\beta\sim (\delta\gamma)^\omega$,
the set $\{(\alpha,(\gamma\delta)^\omega)), (\beta,(\delta\gamma)^\omega),
(A(\gamma\delta)^\omega, B(\delta\gamma)^\omega)\}$ can be easily
checked to be a bisimilar decomposition of $(A\alpha,B\beta)$;
moreover, all strings in the
decomposition have $\bound$-bounded cycles. 
(By our convention $(\delta\gamma)^\omega=\delta$ if
$\|\delta\|=\omega$\,, etc.)

\item
The condition (a) does not hold:

Let us consider a norm-reducing path
$B\gt{a_1}\gamma_1\gt{a_2}\gamma_2\cdots
\gt{a_k}\gamma_k=\varepsilon$ ($k=\|B\|$), and the corresponding path
$B\beta\gt{a_1}\gamma_1\beta\gt{a_2}\gamma_2\beta\cdots
\gt{a_k}\gamma_k\beta=\beta$.
By Prop.~\ref{prop:basicmatching}(2)
there is a path
$A\alpha\gt{a_1}\alpha_1\gt{a_2}\alpha_2\cdots
\gt{a_k}\alpha_k$ such that $\alpha_j\sim \gamma_j\beta$
for $j=1,2,\dots,k$.
Since (a) does not hold, there must be $i\in\{1,2,\dots,k\}$ such that 
$\alpha_i=\alpha$ 
($A$ has been erased, and $\alpha$ has been exposed);
let us put $\gamma=\gamma_i$.
We thus have 
$\alpha\sim \gamma\beta$ where $\|\gamma\|<\|B\|\leq M$.

If $(B\beta)_\prefix\neq\varepsilon$ 
(hence $B$ is the first symbol of the
canonical prefix and $\size(B\beta)=\|B\|+\size(\beta)$) then 
$\size(\gamma\beta)\leq \|\gamma\|+\size(\beta)<\size(B\beta)$.
If $(B\beta)_\prefix=\varepsilon$ 
(hence $B\beta =((B\beta)_\cycle)^\omega$ and
$\size(B\beta)=\size(\beta)\leq\bound$)
then 
$\size(\gamma\beta)\leq \|\gamma\|+\size(\beta)
< M+\bound$.
The assumption
$\size((A\alpha)_\prefix,(B\beta)_\prefix)>2M+\bound$ thus implies
\begin{center}
$\size(\alpha,\gamma\beta)<\size(A\alpha,B\beta)$.
\end{center}
We now explore the following two subcases separately.

\begin{enumerate}
\item
$A\gamma\sim B$:

Here $\{\,(A\gamma,B), (\alpha,\gamma\beta)\,\}$ is a bisimilar
decomposition of $(A\alpha,B\beta)$ (we recall Example~\ref{E:3.8}),
where all strings have $\bound$-bounded cycles.

\item
$A\gamma\not\sim B$ (but $A\gamma\beta\sim B\beta$, since $A\alpha\sim
B\beta$ and $\alpha\sim\gamma\beta$):

Here we use Lemma~\ref{lem:yielddelta}: by putting 
there $\alpha_1=A\gamma$, $\alpha_2=B$ we get $\beta\sim \delta^\omega$
where
$\size(\delta)\leq (2M + \card{\calN}^2\cdot\Mrhs + \Srhs)\cdot
(1+\Srhs)=\bound$.
Hence $\{ (\alpha,\gamma\beta), (\beta,\delta^\omega),
(A\gamma\delta^\omega, B\delta^\omega)\}$ is a bisimilar decomposition
of $(A\alpha, B\beta)$,
where all strings have $\bound$-bounded cycles; since
$\size(A\gamma\delta^\omega,B\delta^\omega)\leq
2M+\bound<\size(A\alpha,B\beta)$, the size conditions
indeed hold.\qed
\end{enumerate}
\end{enumerate}
\end{enumerate}

\begin{lem}\label{lem:completeness}
(Completeness.) There is a polynomial $pol$, used in Point 2 of the
Prover-Refuter game, such that 
$X\sim Y$ implies that Prover has a strategy avoiding
Refuter's win (the play may be infinite).
\end{lem}

\proof
Starting with $X\sim Y$, we let Prover
maintain bisimilarity of (the strings in) each current pair.
In Point 3(b) of the game
Prover only uses 
 bisimilar decompositions of the form   
presented in the case analysis in the proof of
Lemma~\ref{lem:detailedcompleteness}, whenever 
the canonical prefix of a string in the current pair
is bigger than $2M+\bound$.
Doing this, Prover keeps the property that 
the strings in any current pair have $\bound$-bounded cycles.
In Point 3(c) Prover always chooses so that the next current pair is
again bisimilar; Prop.~\ref{prop:invarstep} implies 
that the $\bound$-boundedness of the cycles
is kept.

Adhering to the above strategy, Prover maintains the property 
that the current pair fits into space $2\cdot(2\cdot M+2\cdot\bound+\Srhs)$.
The case analysis in the proof of 
Lemma~\ref{lem:detailedcompleteness} also makes clear 
that
the space $d\cdot \bound$, for a fixed (small) constant $d\in\Nat$
independent of $\calG, X,Y$,
is sufficient for presenting the appropriate decompositions together with the
least congruence proofs.
The claim of the lemma thus easily follows.
\qed

\subsection{Proof of 
Lemma~\ref{lem:yielddelta}}\label{subsec:normedbounds}

We now prove Lemma~\ref{lem:yielddelta}, by which 
a proof of Theorem~\ref{th:bpaintwoexptime}
will be finished.
We assume a BPA system
$\calG=(\calN,\act,\calR)$, with
the associated labelled transition system 
$\ltsact=(\states,\act,(\gt{a})_{a\in\act})$
and with the values
$M,\Mrhs,\Srhs$ (recall Def.~\ref{def:basicsizes} and Prop.~\ref{prop:norms}).
The assumed $\calG$ is general, 
the special case of normed BPA systems
is discussed in the next section.
We first note the following simple fact.

\begin{prop}\label{prop:simpledelta}
If $\sigma\beta\sim\sigma'\beta$ and $\|\sigma\|<\|\sigma'\|$ then
there is $\delta\neq\varepsilon$ such that $\beta\sim\delta\beta$ and
\begin{center}
$\size(\delta)\leq \size(\sigma,\sigma')\cdot(1+\Srhs)$.
\end{center}
\end{prop}

\proof
Suppose $\sigma\beta\sim\sigma'\beta$ and $\|\sigma\|<\|\sigma'\|$;
let $\sigma\gt{v}\varepsilon$ be 
a norm-reducing path. The path $\sigma\beta\gt{v}\beta$ must have
a matching path $\sigma'\beta\gt{v}\tau$ such that
$\beta\sim\tau$ (recall Prop.~\ref{prop:basicmatching}(2)).
Since $\|\sigma'\|>\|\sigma\|$, we can write $\tau=\delta\beta$ where
$\sigma'\gt{v}\delta$ and $\delta\neq\varepsilon$;
we note that $\size(\delta)\leq \size(\sigma')+|v|\cdot\Srhs$
(using Prop.~\ref{prop:invarstep} generously).
Since $|v|=\|\sigma\|\leq \size(\sigma,\sigma')$,
we get
\begin{center}
$\size(\delta)\leq
\size(\sigma,\sigma')+\size(\sigma,\sigma')\cdot\Srhs=
\size(\sigma,\sigma')\cdot(1+\Srhs)$.
\end{center}
\qed

\setcounter{thm}{10}

\begin{lem}
(Repeated.)
If $\alpha_1\not\sim\alpha_2$
and $\alpha_1\beta\sim\alpha_2\beta$ 
then there is $\delta\neq\varepsilon$ such that
$\beta\sim\delta\beta$ 
(and thus $\beta\sim\delta^\omega$)
and
$\size(\delta)\leq (\size(\alpha_1,\alpha_2)+
\card{\calN}^2\cdot\Mrhs+\Srhs)\cdot (1+\Srhs)$.
\end{lem}

\proof
In the assumed BPA system $\calG=(\calN,\act,\calR)$, for each pair
$(A_1,A_2)$ of
nonterminals where $\|A_1\|\leq\|A_2\|<\omega$ 
we fix a norm-reducing path $A_2\gt{u}\gamma$ such that
$\|\gamma\|=\|A_2\|-\|A_1\|$ (hence $|u|=\|A_1\|$).

Now we consider $\alpha_1,\alpha_2,\beta$ such that
$\alpha_1\not\sim\alpha_2$
and $\alpha_1\beta\sim\alpha_2\beta$. 
At least one of $\alpha_1,\alpha_2$ must be normed 
(otherwise $\alpha_1\beta\sim\alpha_1$ and
$\alpha_2\beta\sim\alpha_2$), and we thus
have 
$\|\alpha_1\|\neq \|\alpha_2\|$ or $\|\alpha_1\|=\|\alpha_2\|<\omega$.
If $\|\alpha_1\|\neq \|\alpha_2\|$ then the claim of the lemma
is  true by Prop.~\ref{prop:simpledelta}.
We thus assume $\|\alpha_1\|=\|\alpha_2\|<\omega$, and
imagine a stepwise (not necessarily effective) construction
of a certain sequence 
\begin{equation}\label{eq:decrseq}
(\rho_1,\rho'_1,\mu_1), (\rho_2,\rho'_2,\mu_2), \dots,
(\rho_m,\rho'_m,\mu_m)
\end{equation}
where $(\rho_1,\rho'_1,\mu_1)=(\alpha_1,\alpha_2,\varepsilon)$.
The construction will guarantee that for all $i\in\{1,2,\dots,m\}$
we have $\rho_i\not\sim \rho'_i$, $\mu_i$ is normed,
and $\rho_i\mu_i\beta\sim\rho'_i\mu_i\beta$;
for $i=1$ this holds by the assumptions.
Moreover, we will have $\|\rho_i\|=\|\rho'_i\|<\omega$ for
$i=1,2,\dots,m{-}1$, and  $\|\rho_m\|\neq\|\rho'_m\|$.

Suppose we have constructed 
$(\rho_i,\rho'_i,\mu_i)$ where
$\|\rho_i\|=\|\rho'_i\|<\omega$, $\rho_i\not\sim\rho'_i$,
and $\rho_i\mu_i\beta\sim\rho'_i\mu_i\beta$.
Since both $\rho_i,\rho'_i$ are thus nonempty,
we can write 
\begin{equation}\label{eq:rhoi}
\rho_i=A_1\delta_1,
\rho'_i=A_2\delta_2
\end{equation}
where $A_1,A_2\in\calN$ (and $\|A_1\delta_1\|=
\|A_2\delta_2\|<\omega$).
We assume
$\|A_1\|\leq \|A_2\|$ (otherwise we just swap $\rho_i,\rho'_i$);
let $A_2\gt{u}\gamma$ be the norm-reducing path
which we fixed for $(A_1,A_2)$ above. 
Recall that $|u|=\|A_1\|$, $\|A_1\gamma\|=\|A_2\|$ and note that
$\|\delta_1\|=\|\gamma\delta_2\|$.
We thus have
\begin{equation}\label{eq:auxone}
A_1\delta_1\not\sim A_2\delta_2
\textnormal{ and }
A_1\delta_1\mu_i\beta\sim A_2\delta_2\mu_i\beta
\end{equation}
and we now describe how to choose 
$(\rho_{i+1},\rho'_{i+1},\mu_{i+1})$, 
depending on the following
cases.

\medskip

\begin{enumerate}[(1)]
\item
$(\rho_i,\rho'_i)=(A_1\gamma,A_2)$, i.e. $\delta_1=\gamma$
and
$\delta_2=\varepsilon$ in~(\ref{eq:rhoi}):

\medskip

\noindent
Hence $A_1\gamma\not\sim A_2$
and $A_1\gamma\mu_i\beta\sim A_2\mu_i\beta$.
We fix a 
rule $A_j\gt{a}\sigma_j$, $j\in\{1,2\}$, such that 
for any rule  $A_{3-j}\gt{a}\sigma_{3-j}$ we get
 $\eqlevel(A_1\gamma,A_2)>
 \eqlevel(\sigma_1\gamma,\sigma_2)$;
such $A_j\gt{a}\sigma_j$ exists by Prop.~\ref{prop:basicmatching}(1).
Now we fix 
a rule  $A_{3-j}\gt{a}\sigma_{3-j}$ such that 
$\sigma_1\gamma\mu_i\beta\sim \sigma_2\mu_i\beta$;
such $A_{3-j}\gt{a}\sigma_{3-j}$ exists 
by Prop.~\ref{prop:basicmatching}(2).
Using the fixed rules
 $A_{1}\gt{a}\sigma_{1}$, $A_{2}\gt{a}\sigma_{2}$, we put
\begin{center}
$(\rho_{i+1},\rho'_{i+1},\mu_{i+1})=
(\sigma_1\gamma,\sigma_2,\mu_i)$.
\end{center}
We note the following properties of our choice:
\begin{iteMize}{$\bullet$}
\item
$\eqlevel(\rho_{i+1},\rho'_{i+1})<
\eqlevel(\rho_{i},\rho'_{i})$, 
\item
$\rho_{i+1}\mu_{i+1}\beta\sim\rho'_{i+1}\mu_{i+1}\beta$, 
\item
$\min\,\{\,\|\rho_{i+1}\mu_{i+1}\|,\|\rho'_{i+1}\mu_{i+1}\|\,\}
\leq \|\rho_{i}\mu_{i}\|+\Mrhs-1$
\\
(we cannot have 
$\|\sigma_1\|=\|\sigma_2\|=\omega$ since 
$\sigma_1\gamma\not\sim\sigma_2$ and
$\sigma_1\gamma\mu_i\beta\sim\sigma_2\mu_i\beta$),
\item
$\size(\rho_{i+1}\mu_{i+1},\rho'_{i+1}\mu_{i+1})\leq 
\max\,\{\, \|\rho_i\mu_i\|+\Mrhs-1\,,\, \Srhs \,\}$.
\end{iteMize}
We need to count with $\Srhs$ since one of
$\sigma_{1}$, $\sigma_2$ can be unnormed;
in this case one of $\rho_{i+1}\mu_{i+1},\rho'_{i+1}\mu_{i+1}$
is unnormed and its size is at most $\Srhs$
(using our convention that $\sigma\tau=\sigma$ when
$\|\sigma\|=\omega$).
We have the following two possibilities.
\begin{enumerate}[(a)]
\item
If $\|\sigma_1\gamma\|\neq\|\sigma_2\|$ then
$\|\rho_{i+1}\|\neq\|\rho'_{i+1}\|$ and 
the sequence~(\ref{eq:decrseq}) is completed, i.e.  $i{+}1=m$.
\item
If $\|\sigma_1\gamma\|=\|\sigma_2\|$ then
$\|\rho_{i+1}\mu_{i+1}\|=\|\rho'_{i+1}\mu_{i+1}\|<\omega$.
\end{enumerate}

\medskip

\item
$(\rho_i,\rho'_i)=(A_1\delta_1,A_2\delta_2)\neq(A_1\gamma,A_2)$, 
and we have
\begin{equation}\label{eq:twoconditions}
\eqlevel(A_1\gamma,A_2)\leq \eqlevel(A_1\delta_1,A_2\delta_2)
\textnormal{ and } A_1\gamma\delta_2\mu_i\beta\sim A_2\delta_2\mu_i\beta:
\end{equation}

\noindent
Here we put
\begin{center}
$(\rho_{i+1},\rho'_{i+1},\mu_{i+1})=
(A_1\gamma,A_2,\delta_2\mu_i)$;
\end{center}
this choice has the following properties:
\begin{iteMize}{$\bullet$}
\item
$\eqlevel(\rho_{i+1},\rho'_{i+1})\leq
\eqlevel(\rho_{i},\rho'_{i})$, 
\item
$\rho_{i+1}\mu_{i+1}\beta\sim\rho'_{i+1}\mu_{i+1}\beta$, 
\item
$\|\rho_{i+1}\mu_{i+1}\|=\|\rho'_{i+1}\mu_{i+1}\|
= \|\rho_{i}\mu_{i}\|=\|\rho'_{i}\mu_{i}\|$.
\end{iteMize}
Moreover, for $i{+}1$ the above case (1) will apply.

\medskip

\item
None of (1), (2) applies:

\medskip

\noindent
Since (1) and (2) cover precisely the cases where
the conjunction~(\ref{eq:twoconditions}) holds, 
here we handle the cases where the conjunction~(\ref{eq:twoconditions}) does not hold.
We partition these cases into the disjoint parts (a) and (b) below.

\medskip

\begin{enumerate}
\item
(\ref{eq:twoconditions}) does not hold, and
$\delta_1\mu_i\beta\not\sim \gamma\delta_2\mu_i\beta$:

\medskip

(The reasoning here is based on the fact 
$\delta_1\mu_i\beta\not\sim \gamma\delta_2\mu_i\beta$,
and it could be applied even if (\ref{eq:twoconditions}) would hold.)
\\
We recall $A_1\delta_1\mu_i\beta\sim
A_2\delta_2\mu_i\beta$ from~(\ref{eq:auxone}).
Hence 
the path $A_2\delta_2\mu_i\beta\gt{u}\gamma\delta_2\mu_i\beta$
(corresponding to the fixed norm-reducing path $A_2\gt{u}\gamma$) 
has a matching path  $A_1\delta_1\mu_i\beta\gt{u}$ 
as claimed in Prop.~\ref{prop:basicmatching}(2);
this path cannot finish 
in $\delta_1\mu_i\beta$, since 
$\delta_1\mu_i\beta\not\sim \gamma\delta_2\mu_i\beta$
(i.e., the respective path $A_1\gt{u}$ cannot be norm-reducing).
Though we start with the same norms $\|A_1\delta_1\|=\|A_2\delta_2\|$,
we thus must get a difference of norms in the following sense:
the 
path $A_2\gt{u}\gamma$ has a prefix 
$A_2\gt{u_1}\sigma_2\gt{a}\tau_2$, where $a\in\act$ (and $u_1$ might
be empty),
such that there is a path
$A_1\gt{u_1}\sigma_1\gt{a}\tau_1$ where
$\sigma_1\delta_1\mu_i\beta\sim \sigma_2\delta_2\mu_i\beta$,
$\|\sigma_1\delta_1\|=\|\sigma_2\delta_2\|$,
and 
$\tau_1\delta_1\mu_i\beta\sim \tau_2\delta_2\mu_i\beta$,
$\|\tau_1\delta_1\|>\|\tau_2\delta_2\|$.
Here we put
\begin{center}
$(\rho_{i+1},\rho'_{i+1},\mu_{i+1})=
(\tau_1\delta_1,\tau_2\delta_2,\mu_i)$.
\end{center}
In this case
$\|\rho_{i+1}\|\neq\|\rho'_{i+1}\|$, and~(\ref{eq:decrseq})
is completed, i.e. $i{+}1=m$.
\\
Here we do not claim that $\eqlevel(\rho_{i+1},\rho'_{i+1})\leq
\eqlevel(\rho_{i},\rho'_{i})$ but 
we note the following properties:
\begin{iteMize}{$\bullet$}
\item
$\rho_{i+1}\mu_{i+1}\beta\sim\rho'_{i+1}\mu_{i+1}\beta$, 
\item
$\|\rho_{i+1}\mu_{i+1}\|\neq\|\rho'_{i+1}\mu_{i+1}\|$,
\item
$\min\,\{\,\|\rho_{i+1}\mu_{i+1}\|,\|\rho'_{i+1}\mu_{i+1}\|\,\}
= \|\rho'_{i+1}\mu_{i+1}\| < \|\rho_{i}\mu_{i}\|$,
\item
$\size(\rho_{i+1}\mu_{i+1})\leq 
\max\,\{\, \|\rho_i\mu_i\|+\Mrhs-1\,,\, \Srhs \,\}$.
\end{iteMize}
The last two points follow from the facts that 
$\|\tau_2\delta_2\|< \|A_2\delta_2\|$
(since $A_2\gt{u}\gamma$ is norm-reducing) and that
$\tau_1$ arises by applying a rule to 
$\sigma_1$; thus 
$\|\tau_1\delta_1\|\leq \|\sigma_1\delta_1\|+\Mrhs-1\leq
\|A_1\delta_1\|+\Mrhs-1$ if $\tau_1$ is normed and
$\size(\tau_1)\leq\Srhs$ if $\tau_1$ is unnormed
(in which case $\rho_{i+1}\mu_{i+1}=\tau_1$).

\medskip

\item
(\ref{eq:twoconditions}) does not hold, and
$\delta_1\mu_i\beta\sim \gamma\delta_2\mu_i\beta$:

\medskip

We note that 
$\delta_1\mu_i\beta\sim \gamma\delta_2\mu_i\beta$ implies
$A_1\delta_1\mu_i\beta\sim A_1\gamma\delta_2\mu_i\beta$, 
and the
assumption $A_1\delta_1\mu_i\beta\sim
A_2\delta_2\mu_i\beta$~(\ref{eq:auxone}) then yields
$A_1\gamma\delta_2\mu_i\beta\sim
A_2\delta_2\mu_i\beta$; 
the second conjunct in~(\ref{eq:twoconditions}) thus holds.
Hence the first conjunct does not hold, and we have
\begin{center}
$\eqlevel(A_1\gamma,A_2)>\eqlevel(A_1\delta_1,A_2\delta_2)$.
\end{center}
We thus have
$\eqlevel(A_1\gamma\delta_2,A_2\delta_2)> \eqlevel(A_1\delta_1,A_2\delta_2)$
(by Prop.~\ref{prop:basicmatching}(4)); this implies that
$\eqlevel(A_1\delta_1,A_1\gamma\delta_2)=
\eqlevel(A_1\delta_1,A_2\delta_2)$ 
(by Prop.~\ref{prop:basicmatching}(3)).
\\
Since $\eqlevel(A_1\delta_1,A_1\gamma\delta_2)\geq 
\|A_1\|+\eqlevel(\delta_1,\gamma\delta_2)$
(by Prop.~\ref{prop:simplenorm}(3)), we get
\begin{center}
$\eqlevel(\delta_1,\gamma\delta_2)\leq
\eqlevel(A_1\delta_1,A_1\gamma\delta_2)-\|A_1\|
=
\eqlevel(A_1\delta_1,A_2\delta_2)-\|A_1\|$. 
\end{center}
We put
\begin{center}
$(\rho_{i+1},\rho'_{i+1},\mu_{i+1})=
(\delta_1,\gamma\delta_2,\mu_i)$
\end{center}
and note the following properties:
\begin{iteMize}{$\bullet$}
\item
$\eqlevel(\rho_{i+1},\rho'_{i+1})\leq 
\eqlevel(\rho_{i},\rho'_{i})-\|A_1\|<\eqlevel(\rho_{i},\rho'_{i})$,
\item
$\rho_{i+1}\mu_{i+1}\beta\sim\rho'_{i+1}\mu_{i+1}\beta$, 
\item
$\|\rho_{i+1}\mu_{i+1}\|=\|\rho'_{i+1}\mu_{i+1}\|=\|\rho_i\mu_i\|-\|A_1\|$.
\end{iteMize}
\end{enumerate}
\end{enumerate}

\medskip

\noindent
If we construct a sequence~(\ref{eq:decrseq}) by performing
the above described step for $i=1,2,3,\dots$,
we obviously maintain the properties
 $\rho_{i}\not\sim\rho'_{i}$
and $\rho_{i}\mu_{i}\beta\sim\rho'_{i}\mu_{i}\beta$.
When some $(\rho_i,\rho'_i,\mu_i)$ where 
$\|\rho_i\|\neq \|\rho'_i\|$ is constructed, the construction
ends ($i=m$ in~(\ref{eq:decrseq})),
and this is the only way how to end.
The end is reached whenever the case (3a) applies; 
another possibility occurs in the case (1).
We also maintain that $\mu_i$ is normed; $\mu_i$ is ``increasing''
in the sense that 
 $\mu_i$ is a suffix of $\mu_{i+1}$ (for $i<m$).

Informally speaking, 
the ``head eq-level'' is decreasing. 
More precisely, if (1), (2), or (3b) applies to $i$ then 
we have 
$\eqlevel(\rho_{i},\rho'_{i})\geq 
\eqlevel(\rho_{i+1},\rho'_{i+1})$; if (3a) applies then we do not care 
since the construction finishes (with $i{+}1=m$).
In (1) and (3b) the head eq-level is even \emph{strictly} decreasing, i.e.
$\eqlevel(\rho_{i},\rho'_{i})> \eqlevel(\rho_{i+1},\rho'_{i+1})$.
We thus cannot use (1)
for the same pair $(A_1,A_2)$ twice;
this implies that (1) cannot be used more than
$\card{\calN}^2$ 
times (which is a generous upper bound).
Since any use of (2) for $i$ entails using (1) for $i{+}1$,
the head eq-level decreasing guarantees that 
the construction must end eventually,
reaching some $(\rho_m,\rho'_m,\mu_m)$ where $\|\rho_m\|\neq
\|\rho'_m\|$.

We recall that
$\min\,\{\,\|\rho_1\mu_1\|,
\|\rho'_1\mu_1\|\,\}=\|\alpha_1\|=\|\alpha_2\|<\omega$.
We can easily check that for each $i\in\{1,2,\dots,m{-}1\}$ we have:
\begin{iteMize}{$\bullet$}
\item
if (1)
applies to $i$ then
$\min\,\{\,\|\rho_{i+1}\mu_{i+1}\|,\|\rho'_{i+1}\mu_{i+1}\|\,\}\leq 
\min\,\{\,\|\rho_{i}\mu_{i}\|,\|\rho'_{i}\mu_{i}\|\,\}+\Mrhs$; 
\item 
if (2) or (3) applies to $i$ then
$\min\,\{\,\|\rho_{i+1}\mu_{i+1}\|,\|\rho'_{i+1}\mu_{i+1}\|\,\}\leq 
\min\,\{\,\|\rho_{i}\mu_{i}\|,\|\rho'_{i}\mu_{i}\|\,\}$.
\end{iteMize}
We thus have
\begin{center}
$\min\{\|\rho_m\mu_m\|, \|\rho'_m\mu_m\|\}\leq
\|\alpha_1\|+\card{\calN}^2\cdot\Mrhs$. 
\end{center}
If both $\rho_m$, $\rho'_m$ are normed then
\begin{center}
$\max\{\|\rho_m\mu_m\|, \|\rho'_m\mu_m\|\}\leq
\|\alpha_1\|+\card{\calN}^2\cdot\Mrhs + \Mrhs$;
\end{center}
in fact, 
$\max\{\|\rho_m\mu_m\|, \|\rho'_m\mu_m\|\}\leq
\min\{\|\rho_m\mu_m\|, \|\rho'_m\mu_m\|\}+\Mrhs$, as can be checked in
(1) and (3a).
If one of $\rho_m$, $\rho'_m$ is unnormed then 
its size 
is at most $\Srhs$.
We can thus safely confirm that 
\begin{center}
$\size(\rho_m\mu_m, \rho'_m\mu_m)\leq 
\size(\alpha_1,\alpha_2)+\card{\calN}^2\cdot\Mrhs+\Srhs$.
\end{center}
Since
$\|\rho_m\mu_m\|\neq \|\rho'_m\mu_m\|$ and
$\rho_m\mu_m\beta\sim\rho'_m\mu_m\beta$,
Prop.~\ref{prop:simpledelta} finishes the proof.
\qed

\section{Exponential bound on eq-levels in normed BPA systems}\label{sec:eqlevelbounds}

A BPA system is normed if each nonterminal is normed:
\begin{defi}
A \emph{BPA system} $\calG=(\calN,\act,\calR)$
is \emph{normed} if
$\|A\|<\omega$ for all $A\in\calN$.
\end{defi}
\noindent
\emph{Convention.}
In this section we stipulate  $\states=\calN^*$
in the LTS $\ltsact=(\states,\act,(\gt{a})_{a\in\act})$; we thus do
not consider infinite regular strings (since they are unnormed).

\medskip

As already mentioned, the problem \textsc{BPA-Bisim}
restricted to normed BPA systems is known to be in $\PTIME$.
Nevertheless it is easy to construct an example where $\eqlevel(X,Y)$
(for $X\not\sim Y$)
is exponential in the size of the given normed BPA system $\calG$;
e.g., in Remark after Prop.~\ref{prop:norms} we 
have $\eqlevel(A_k,A_{k-1})=\|A_{k-1}\|=2^{k-1}-1$.

An exponential upper bound on the eq-levels in the normed case seems 
to be only implicit in the literature;
we thus show a bound explicitly here, as Theorem~\ref{thm:normedbound}.
In principle, we use again the construction from the proof of
Lemma~\ref{lem:yielddelta} in Subsection~\ref{subsec:normedbounds},
but now in a different setting and with a
different aim. It is easy to note that in the normed case we cannot have
$\alpha_1\not\sim\alpha_2$ and $\alpha_1\beta\sim\alpha_2\beta$;
but this is not a problem, we do not need such $\beta$ here.
We will construct a sequence
like~(\ref{eq:decrseq}), with the decreasing
head  eq-levels $\eqlevel(\rho_i,\rho'_i)$, but we will now take also the
``overall'' eq-levels  $\eqlevel(\rho_i\mu_i,\rho'_i\mu_i)$ into
account. These eq-levels were of no interest in 
Subsection~\ref{subsec:normedbounds}
(there we just took care that 
$\eqlevel(\rho_i\mu_i\beta,\rho'_i\mu_i\beta)=\omega$);
here these overall eq-levels add technical
complications since they can evolve differently than the head
eq-levels.
We remove these complications when we arrange that 
$\eqlevel(\rho_i\mu_i,\rho'_i\mu_i)=\eqlevel(\rho_i,\rho'_i)+\|\mu_i\|$;
that's why we introduce the following completion of a normed system
with a special unnormed nonterminal.

\begin{defi}
For a normed BPA system $\calG=(\calN,\act,\calR)$, by the
\emph{completion of} $\calG$ we mean the BPA system
 $\calG'=(\calN\cup\{U\},\act,\calR')$ where 
$U$ is a special (unnormed) nonterminal, and 
$\calR'=\calR\cup \{U\gt{a}U\mid a\in\act\}\cup \{A\gt{a}U\mid
A\in\calN, a\in\act\}$.
\end{defi}

By our conventions, in the LTS 
$\ltsactprime=(\statesprime,\act,(\gt{a})_{a\in\act})$ 
we have $\statesprime=\calN^*\cup\{\alpha U \mid \alpha\in \calN^*\}$.
In $\ltsactprime$ we obviously have $\eqlevel(\alpha,\beta)=0$ iff
precisely one of $\alpha,\beta$ is $\varepsilon$.
Other useful properties of $\ltsactprime$ are captured in
Prop.~\ref{prop:helpltsprime},
but we first make clear 
that an upper bound on eq-levels in
$\ltsactprime$
is also an upper bound on eq-levels in $\ltsact$.

\begin{prop}\label{prop:ltsacttoltsactprime}
\hfill
\begin{enumerate}[\em(1)]
\item
$\eqlevel(\gamma_1,\gamma_2)$ in $\ltsact$ is not bigger 
than $\eqlevel(\gamma_1,\gamma_2)$ in $\ltsactprime$.
\item
In  $\ltsactprime$ we have $\alpha\sim U$ iff $\|\alpha\|=\omega$.
\item
For any $\gamma_1,\gamma_2\in\calN^*$ we have 
$\gamma_1\sim\gamma_2$ in $\ltsact$ iff $\gamma_1\sim\gamma_2$ in
$\ltsactprime$.
\\
(Hence if $\eqlevel(\gamma_1,\gamma_2)$ is finite in $\ltsact$ then it
is finite in  $\ltsactprime$ as well.)
\end{enumerate}
\end{prop}

\proof
(1) If $\gamma_1\sim_i\gamma_2$ in $\ltsact$
then $\gamma_1\sim_i\gamma_2$ in $\ltsactprime$, as can be shown 
by induction on $i$, when
noting that each move $\gamma_j\gt{a}U$ can be matched by $\gamma_{3-j}\gt{a}U$
if $\gamma_{3-j}\neq\varepsilon$.

(2) If $\|\alpha\|<\|\beta\|$ then $\alpha\not\sim \beta$
(recall Prop.~\ref{prop:simplenorm}(1)); on the other hand, 
the set $\{(\alpha,\beta); \|\alpha\|=\|\beta\|=\omega\}$ is here a
bisimulation.

(3) From Point 1 we get that $\gamma_1\sim\gamma_2$ in $\ltsact$
implies  $\gamma_1\sim\gamma_2$ in $\ltsactprime$; on the other hand, 
$\{(\alpha,\beta)\in\calN^*\times\calN^* \mid \alpha\sim\beta$ in
$\ltsactprime\}$ can be easily checked to be a bisimulation in
$\ltsact$.
\qed

\begin{prop}\label{prop:helpltsprime}
In $\ltsactprime$ the following claims hold: 
\begin{enumerate}[\em(1)]
\item
$\eqlevel(\gamma_1,\gamma_2)=0$ iff precisely one of
$\gamma_1,\gamma_2$ is 
the empty word $\varepsilon$. 
\item
$\eqlevel(\gamma_1,\gamma_2)\geq \min\{\|\gamma_1\|,\|\gamma_2\|\}$.
\item
If $\|\gamma_1\|\neq\|\gamma_2\|$ then
$\eqlevel(\gamma_1,\gamma_2)=\min\{\|\gamma_1\|,\|\gamma_2\|\}$.
\item
Suppose $\|\alpha_1\|\leq\|\alpha_2\|<\omega$ and
$\alpha_2\gt{u}\gamma$ is a norm-reducing path where $|u|=\|\alpha_1\|$
(and thus $\|\gamma\|=\|\alpha_2\|-\|\alpha_1\|$).
Then for any $\delta_1,\delta_2$ we have
\\
$\eqlevel(\delta_1,\gamma\delta_2)\geq 
\eqlevel(\alpha_1\delta_1,\alpha_2\delta_2)-\|\alpha_1\|$.
\item
$\eqlevel(\sigma_1\mu,\sigma_2\mu)=\eqlevel(\sigma_1,\sigma_2)+\|\mu\|$.
\end{enumerate}
\end{prop}

\proof
Points 1,2,3 are easy to observe.

\begin{enumerate}[(1)]
 
\item[(4)]If $\|\alpha_1\delta_1\|\neq\|\alpha_2\delta_2\|$ then
$\|\delta_1\|=\|\alpha_1\delta_1\|-\|\alpha_1\|\neq
\|\alpha_2\delta_2\|-\|\alpha_1\|=\|\gamma\delta_2\|$, and
(3) implies 
\begin{center}
$\eqlevel(\delta_1,\gamma\delta_2)=\min\{\|\delta_1\|,\|\gamma\delta_2\|\}=
\min\{\|\alpha_1\delta_1\|,\|\alpha_1\gamma\delta_2\|\}-\|\alpha_1\|=
\min\{\|\alpha_1\delta_1\|,\|\alpha_2\delta_2\|\}-\|\alpha_1\|=
\eqlevel(\alpha_1\delta_1,\alpha_2\delta_2)-\|\alpha_1\|$.
\end{center}
We now assume $\|\alpha_1\delta_1\|=\|\alpha_2\delta_2\|$
(hence also $\|\delta_1\|=\|\gamma\delta_2\|$)
and
we contradict the assumption 
\begin{equation}\label{eq:contreqlevel}
\eqlevel(\delta_1,\gamma\delta_2)< 
\eqlevel(\alpha_1\delta_1,\alpha_2\delta_2)-\|\alpha_1\|
\end{equation}
as follows.
By Prop.~\ref{prop:basicmatching}(2),
the norm-reducing path $\alpha_2\delta_2\gt{u}\gamma\delta_2$ has
a matching path $\alpha_1\delta_1\gt{u}\sigma\delta_1$ 
where $\eqlevel(\sigma\delta_1,\gamma\delta_2)\geq
\eqlevel(\alpha_1\delta_1,\alpha_2\delta_2)-\|\alpha_1\|$.
Hence $\sigma\neq\varepsilon$
(i.e., $\alpha_1\gt{u}\sigma$ is not norm-reducing), and thus
$\|\sigma\delta_1\|>\|\gamma\delta_2\|$, which entails
 $\eqlevel(\sigma\delta_1,\gamma\delta_2)=\|\gamma\delta_2\|$.
 Since $\eqlevel(\delta_1,\gamma\delta_2)\geq\|\gamma\delta_2\|$
(by (2)), by~(\ref{eq:contreqlevel}) we would get a contradiction:
\begin{center}
$\|\gamma\delta_2\|\leq\eqlevel(\delta_1,\gamma\delta_2)< 
\eqlevel(\alpha_1\delta_1,\alpha_2\delta_2)-\|\alpha_1\|
\leq \eqlevel(\sigma\delta_1,\gamma\delta_2)=\|\gamma\delta_2\|$.
\end{center}

\item[(5)]
The equality surely holds if $\|\mu\|=\omega$ 
(in which case $\sigma_1\mu\sim U\sim \sigma_2\mu$)
or if $\sigma_1\sim\sigma_2$; we
thus further assume that $\mu$ is normed and
$\eqlevel(\sigma_1,\sigma_2)<\omega$.
\begin{iteMize}{$\bullet$}
\item
We show
$\eqlevel(\sigma_1\mu,\sigma_2\mu)\leq\eqlevel(\sigma_1,\sigma_2)+\|\mu\|$
by
induction on $\eqlevel(\sigma_1,\sigma_2)$.
\\
If $\eqlevel(\sigma_1,\sigma_2)=0$ then precisely one of
$\sigma_1,\sigma_2$ is
$\varepsilon$, and $\eqlevel(\sigma_1\mu,\sigma_2\mu)=\|\mu\|$.
\\
If $\eqlevel(\sigma_1,\sigma_2)=n{+}1$
(which entails $\sigma_1\neq\varepsilon$, $\sigma_2\neq\varepsilon$)
then 
by Prop.~\ref{prop:basicmatching}(1,2) there are 
some transitions $\sigma_1\gt{a}\tau_1$ and
$\sigma_{2}\gt{a}\tau_{2}$ 
such that
\begin{enumerate}[(1)]
\item
$\eqlevel(\tau_1,\tau_2)<\eqlevel(\sigma_1,\sigma_2)$,
and 
\item
$\eqlevel(\tau_1\mu,\tau_2\mu)\geq
\eqlevel(\sigma_1\mu,\sigma_2\mu)-1$.
\end{enumerate}
Since $\eqlevel(\tau_1\mu,\tau_2\mu)\leq \eqlevel(\tau_1,\tau_2)+\|\mu\|$
by the induction hypothesis, we have
$\eqlevel(\sigma_1\mu,\sigma_2\mu)\leq
1+\eqlevel(\tau_1,\tau_2)+\|\mu\|\leq
\eqlevel(\sigma_1,\sigma_2)+\|\mu\|$.
\item
We show
$\eqlevel(\sigma_1\mu,\sigma_2\mu)\geq\eqlevel(\sigma_1,\sigma_2)+\|\mu\|$
by induction on $\eqlevel(\sigma_1\mu,\sigma_2\mu)$,
excluding the trivial case $\eqlevel(\sigma_1\mu,\sigma_2\mu)=\omega$.
\\
The case $\eqlevel(\sigma_1\mu,\sigma_2\mu)=0$ is trivial
since it entails $\mu=\varepsilon$.
\\
If $\eqlevel(\sigma_1\mu,\sigma_2\mu)=n{+}1$ 
then at most one of $\sigma_1,\sigma_2$ can be empty.
If we have $\sigma_j=\varepsilon$ ($j\in\{1,2\}$) then
$\eqlevel(\sigma_1,\sigma_2)=0$ and 
$\eqlevel(\sigma_1\mu,\sigma_2\mu)=\|\mu\|$ (by (3)); the claim thus holds.
If both $\sigma_1,\sigma_2$ are nonempty then
by Prop.~\ref{prop:basicmatching}(1,2) there are 
some transitions $\sigma_1\gt{a}\tau_1$ and
$\sigma_{2}\gt{a}\tau_{2}$ 
such that
\begin{enumerate}[(1)]
\item
$\eqlevel(\tau_1\mu,\tau_2\mu)<
\eqlevel(\sigma_1\mu,\sigma_2\mu)$, and
\item
$\eqlevel(\tau_1,\tau_2)\geq \eqlevel(\sigma_1,\sigma_2)-1$.
\end{enumerate}
Since $\eqlevel(\tau_1\mu,\tau_2\mu)\geq\eqlevel(\tau_1,\tau_2)+\|\mu\|$
by the induction hypothesis, we have
 $\eqlevel(\sigma_1\mu,\sigma_2\mu)\geq 1+\eqlevel(\tau_1,\tau_2)+\|\mu\|
 \geq \eqlevel(\sigma_1,\sigma_2)+\|\mu\|$.\qed\smallskip
\end{iteMize}
\end{enumerate}

\noindent We now prove the announced theorem.  Let us recall that the value
$\Mrhs$ (in Def.~\ref{def:basicsizes}) is bounded by an exponential
function of the size of $\calG$ (by Prop.~\ref{prop:norms}).

\begin{thm}\label{thm:normedbound}
Let $\calG=(\calN,\act,\calR)$ be
a normed BPA system, 
and $\Mrhs=\max\,\{\|\alpha\|;$ there is a rule $A\gt{a}\alpha$ in
$\calR\,\}$.
If $\alpha_1\not\sim\alpha_2$ then 
$\eqlevel(\alpha_1,\alpha_2)\leq 
\min\{\|\alpha_1\|,\|\alpha_2\|\}+\card{\calN}^2\cdot\Mrhs$.
\end{thm}

\proof
If $\|\alpha_1\|<\|\alpha_2\|$ then
$\eqlevel(\alpha_1,\alpha_2)\leq \|\alpha_1\|$,
as we noted in Prop.~\ref{prop:simplenorm}(1) for general BPA systems.
We thus consider $\alpha_1\not\sim\alpha_2$ where
$\|\alpha_1\|=\|\alpha_2\|$, and we will work
in the LTS $\ltsactprime$, where $\calG'$ is the completion
of $\calG$; the achieved upper bound will be also valid for
$\ltsact$ by Prop.~\ref{prop:ltsacttoltsactprime}(1,3).
We will construct a sequence 
\begin{equation}\label{eq:decrseqtwo}
(\rho_1,\rho'_1,\mu_1), (\rho_2,\rho'_2,\mu_2), \dots,
(\rho_m,\rho'_m,\mu_m)
\end{equation}
where $(\rho_1,\rho'_1,\mu_1)=(\alpha_1,\alpha_2,\varepsilon)$.
We use a slightly modified process of constructing 
the sequence~(\ref{eq:decrseq}) 
in the proof of Lemma~\ref{lem:yielddelta} in 
Subsection~\ref{subsec:normedbounds}. 
Given  $(\rho_i,\rho'_i,\mu_i)$, where $\|\rho_i\|=\|\rho'_i\|<\omega$
and
$\eqlevel(\rho_i\mu_i,\rho'_i\mu_i)=\eqlevel(\rho_i,\rho'_i)+\|\mu\|<\omega$,
we now 
construct $(\rho_{i+1},\rho'_{i+1},\mu_{i+1})$.
As in  the proof of Lemma~\ref{lem:yielddelta}, we write
\begin{equation}\label{eq:rhoitwo}
\rho_i=A_1\delta_1,
\rho'_i=A_2\delta_2
\end{equation}
where 
$\|A_1\|\leq \|A_2\|$ and 
we assume that  the pair $(A_1,A_2)$ has 
a fixed  norm-reducing path
$A_2\gt{u}\gamma$ such that 
$\|A_1\gamma\|=\|A_2\|$; we thus also have
$\|\delta_1\|=\|\gamma\delta_2\|$.

\medskip

\begin{enumerate}[(1)]
\item
$(\rho_i,\rho'_i)=(A_1\gamma,A_2)$, i.e. $\delta_1=\gamma$
and
$\delta_2=\varepsilon$ in~(\ref{eq:rhoitwo}):

\medskip

\noindent
By Prop.~\ref{prop:basicmatching}(1,2) there are rules
$A_{1}\gt{a}\sigma_{1}$, $A_{2}\gt{a}\sigma_{2}$
such that $\eqlevel(\sigma_1\gamma,\sigma_2)=
\eqlevel(A_1\gamma,A_2)-1$
(and where we thus do not have
$\sigma_1=\sigma_2=U$).
We put 
\begin{center}
$(\rho_{i+1},\rho'_{i+1},\mu_{i+1})=
(\sigma_1\gamma,\sigma_2,\mu_i)$,
\end{center}
and we note (by recalling that 
$\eqlevel(\rho\mu,\rho'\mu)=\eqlevel(\rho,\rho')+\|\mu\|$):
\begin{iteMize}{$\bullet$}
\item
$\eqlevel(\rho_{i+1},\rho'_{i+1})=
\eqlevel(\rho_{i},\rho'_{i})-1$, 
\item
$\eqlevel(\rho_{i+1}\mu_{i+1},\rho'_{i+1}\mu_{i+1})=
\eqlevel(\rho_{i}\mu_i,\rho'_{i}\mu_i)-1$, 
\item
$\min\,\{\,\|\rho_{i+1}\mu_{i+1}\|,\|\rho'_{i+1}\mu_{i+1}\|\,\}
\leq \|\rho_{i}\mu_{i}\|+\Mrhs-1$.
\end{iteMize}
We have the following two possibilities.
\begin{enumerate}
\item
If $\|\sigma_1\gamma\|\neq\|\sigma_2\|$ then
$\|\rho_{i+1}\|\neq\|\rho'_{i+1}\|$ and 
the sequence~(\ref{eq:decrseqtwo}) is completed,
i.e.  
$i{+}1=m$. In this case
\begin{center}
$\eqlevel(\rho_m\mu_m,\rho'_m\mu_m)=
\min\,\{\,\|\rho_{m}\mu_{m}\|,\|\rho'_{m}\mu_{m}\|\,\}$.
\end{center}
\item
If $\|\sigma_1\gamma\|=\|\sigma_2\|$ then
$\|\rho_{i+1}\mu_{i+1}\|=\|\rho'_{i+1}\mu_{i+1}\|<\omega$.
\end{enumerate}

\medskip

\item
$(\rho_i,\rho'_i)=(A_1\delta_1,A_2\delta_2)\neq(A_1\gamma,A_2)$
and 
$\eqlevel(A_1\gamma\delta_2,A_2\delta_2)=\eqlevel(A_1\delta_1,A_2\delta_2)$:

\medskip

\noindent
We put
\begin{center}
$(\rho_{i+1},\rho'_{i+1},\mu_{i+1})=
(A_1\gamma,A_2,\delta_2\mu_i)$,
\end{center}
and note:
\begin{iteMize}{$\bullet$}
\item
$\eqlevel(\rho_{i+1},\rho'_{i+1})=
\eqlevel(\rho_{i},\rho'_{i})-\|\delta_2\|\leq\eqlevel(\rho_{i},\rho'_{i})$, 
\item
$\eqlevel(\rho_{i+1}\mu_{i+1},\rho'_{i+1}\mu_{i+1})=
\eqlevel(\rho_{i}\mu_i,\rho'_{i}\mu_i)$,
\item
$\|\rho_{i+1}\mu_{i+1}\|=\|\rho'_{i+1}\mu_{i+1}\|
= \|\rho_{i}\mu_{i}\|=\|\rho'_{i}\mu_{i}\|$.
\end{iteMize}
Moreover, for $i{+}1$ the above case (1) will apply.

\medskip

\item
$\eqlevel(A_1\gamma\delta_2,A_2\delta_2)\neq\eqlevel(A_1\delta_1,A_2\delta_2)$
(which entails $(\rho_i,\rho'_i)\neq(A_1\gamma,A_2)$):

\medskip

\noindent
We thus have $\eqlevel(A_1\delta_1,A_1\gamma\delta_2)\leq
\eqlevel(A_1\delta_1,A_2\delta_2)$,
by
Prop.~\ref{prop:basicmatching}(3).
\\
Since $\eqlevel(A_1\delta_1,A_1\gamma\delta_2)\geq \|A_1\|+
\eqlevel(\delta_1,\gamma\delta_2)$ (by
Prop.~\ref{prop:simplenorm}(3)), we get
\begin{center}
$\eqlevel(\delta_1,\gamma\delta_2)\leq \eqlevel(A_1\delta_1,A_1\gamma\delta_2)
-\|A_1\|\leq 
\eqlevel(A_1\delta_1,A_2\delta_2)-\|A_1\|$.
\end{center}
On the other hand, Prop.~\ref{prop:helpltsprime}(4) implies 
\begin{center}
$\eqlevel(\delta_1,\gamma\delta_2)\geq
\eqlevel(A_1\delta_1,A_2\delta_2)-\|A_1\|$.
\end{center}
Hence $\eqlevel(\delta_1,\gamma\delta_2)=
\eqlevel(A_1\delta_1,A_2\delta_2)-\|A_1\|$. 
We put
\begin{center}
$(\rho_{i+1},\rho'_{i+1},\mu_{i+1})=
(\delta_1,\gamma\delta_2,\mu_i)$,
\end{center}
and note:
\begin{iteMize}{$\bullet$}
\item
$\eqlevel(\rho_{i+1},\rho'_{i+1})=
\eqlevel(\rho_{i},\rho'_{i})-\|A_1\|$, 
\item
$\eqlevel(\rho_{i+1}\mu_{i+1},\rho'_{i+1}\mu_{i+1})=
\eqlevel(\rho_{i}\mu_i,\rho'_{i}\mu_i)-\|A_1\|$,
\item
$\|\rho_{i+1}\mu_{i+1}\|=\|\rho'_{i+1}\mu_{i+1}\|
= \|\rho_{i}\mu_{i}\|-\|A_1\|$.
\end{iteMize}
\end{enumerate}

\medskip

\noindent
As in Subsection~\ref{subsec:normedbounds}, 
due to eq-level decreasing
the case (1) cannot 
apply more than $\card{\calN}^2$ times, and the construction must end
eventually, with $\|\rho_m\mu_m\|\neq\|\rho'_m\mu_m\|$ arising in (1a).
Let us now put
\begin{center}
$e_i=\eqlevel(\rho_i\mu_i,\rho'_i\mu_i)$, and
$d_i=e_i-\min\,\{\,\|\rho_{i}\mu_{i}\|,\|\rho'_{i}\mu_{i}\|\,\}$.
\end{center}
In fact, in (1a) we noted that $d_m=0$.
If (2) or (3) applies to $i$ then we obviously have $d_i=d_{i+1}$.
We can also easily check that 
if (1) applies to $i$ then
\begin{center}
$e_{i+1}-\min\,\{\,\|\rho_{i+1}\mu_{i+1}\|,\|\rho'_{i+1}\mu_{i+1}\|\,\}
\geq  
(e_{i}-1)-(\min\,\{\,\|\rho_{i}\mu_{i}\|,\|\rho'_{i}\mu_{i}\|\,\}+\Mrhs-1)$.
\end{center}
This yields $d_{i+1}\geq d_i - \Mrhs$, hence $d_i\leq d_{i+1}+\Mrhs$.
We thus deduce $d_1\leq\card{\calN}^2\cdot\Mrhs$, i.e.,
$e_1\leq \min\,\{\,\|\rho_{1}\mu_{1}\|,\|\rho'_{1}\mu_{1}\|\,\}+
\card{\calN}^2\cdot\Mrhs$.
Since 
$(\rho_1,\rho'_1,\mu_1)=(\alpha_1,\alpha_2,\varepsilon)$, we get
\[\eqlevel(\alpha_1,\alpha_2)\leq
\min\,\{\,\|\alpha_1\|,\|\alpha_2\|\,\}+\card{\calN}^2\cdot\Mrhs\,.
\eqno{\qEd}
\]

\section{Additional remarks}\label{sec:addition}

\noindent
Lemma~\ref{lem:detailedcompleteness} shows that
the pairs $(\alpha,\beta)$ 
where $\alpha\sim\beta$, $\alpha,\beta$ have $\calE$-bounded cycles,
and $\size(\alpha_\prefix,\beta_\prefix)\leq 2M+\calE$ 
create
a \emph{basis} for $\calG$, 
similar to the bisimulation base 
of~\cite{DBLP:conf/mfcs/BurkartCS95} but with explicit regular
strings. 
We could construct the basis by a standard 
coinductive approach (building a
sequence of decreasing overapproximations).
Each of the pairs in the basis fits into exponential space,
and their number is thus at most 
double exponential.

Among the related topics for future research, the obvious one is 
the question how to close the gap  between \ExpTime and
2-\ExpTime for bisimilarity on BPA.
Other examples of research topics follow from the fact that 
 BPA processes can be viewed as being generated by pushdown automata
with a single control state and no $\varepsilon$-transitions.
S\'enizergues~\cite{Seni05} showed  the decidability of bisimilarity 
for general pushdown processes where $\varepsilon$-transitions are
deterministic and popping;
it seems interesting to explore the decomposition approach here as
well, using \emph{regular terms} (as in~\cite{JancarLICS12}). 
One indication that this more general problem is also more complicated 
is a recent announcement~\cite{BGKM12} that 
its computational complexity is nonelementary.
We can also mention that bisimilarity of pushdown processes with
\emph{nondeterministic} popping $\varepsilon$-transitions is 
undecidable~\cite{DBLP:journals/jacm/JancarS08}; this was shown by
using so called ``Defender's Forcing'', which was recently also used
to show undecidability for $2^{nd}$-order pushdown processes with no
$\varepsilon$-transitions~\cite{DBLP:conf/fsttcs/BroadbentG12}.
The decidability question for BPA with $\varepsilon$-transitions
(i.e., the weak bisimilarity problem for BPA) is still open.

\subsection*{Acknowledgement}

The author cordially thanks to anonymous reviewers for helpful comments and
suggestions.

\bibliographystyle{abbrv}
\bibliography{root}

\end{document}